\documentclass[a4paper,11pt]{article}
\pdfoutput=1 % if your are submitting a pdflatex (i.e. if you have
\usepackage{jinstpub} % for details on the use of the package, please
\usepackage{stackengine}
\usepackage{upgreek}
\usepackage{threeparttable}
\usepackage{cases}
\usepackage{adjustbox}
\usepackage{rotating}
\usepackage{lscape}
\usepackage{afterpage}

\usepackage{amsmath}
\usepackage{upgreek}
\usepackage{cases}
\usepackage{fixltx2e}
\usepackage{threeparttable}
\usepackage{hyperref}
\usepackage{subfig}
\usepackage{graphicx}

\title{Measurements and analysis of different front-end configurations for monolithic SiGe BiCMOS pixel detectors for HEP applications}

\author[a,b,1]{F. Martinelli\note{Corresponding author.}}
\author[c]{, C. Magliocca}
\author[c]{, R. Cardella}
\author[b]{, E. Charbon}
\author[c]{, G. Iacobucci}
\author[a,c]{, M. Nessi}
\author[a,c]{, L. Paolozzi}
\author[d]{, H. Rücker}
\author[c]{and P. Valerio}

\affiliation[a]{The European Organization for Nuclear Research (CERN), \\Espl. des Particules 1, 1211 Meyrin, Switzerland}
\affiliation[b]{École polytechnique fédérale de Lausanne (EPFL), Advanced Quantum Architecture (AQUA) Laboratory\\Rue de la Maladière 71C, 2002 Neuchâtel, Switzerland}
\affiliation[c]{University of Geneva, Département de physique nucléaire et corpusculaire (DPNC)\\Quai Ernest-Ansermet 24, 1205 Geneva, Switzerland}
\affiliation[d]{Innovations for High Performance (IHP) Microelectronics,\\Im Technologiepark 25, 15236 Frankfurt (Oder), Germany}

\emailAdd{fulvio.martinelli@cern.ch}

\abstract{This paper presents a small-area monolithic pixel detector  ASIC designed in 130 nm SiGe BiCMOS technology for the upgrade of the pre-shower detector of the FASER experiment at CERN. The purpose of this prototype  is to study the integration of fast front-end electronics inside the sensitive area of the pixels and to identify the  configuration that could satisfy at best the specifications of the experiment. Self-induced noise, instabilities and cross-talk were minimised to cope with the several challenges  associated to the integration of pre-amplifiers and discriminators inside the pixels. 
The methodology used in the characterisation and   the design choices will also be described.
Two of the variants studied here will be implemented in the pre-production ASIC of the FASER experiment pre-shower for further tests. }


\begin{document}
\maketitle
\flushbottom

\section{Silicon pixel detectors for electromagnetic shower reconstruction}
\label{sec:intr}
%focalizzarsi su monolithic pixel detectors (vedi da tesi)
%integrazione con l'elettronica ancje per HEP
%monolithic BiCMOS con elettronica integrata -> FASER uno dei primi esperimenti
%introduci velocissimo FASER. Non troppo 
%metti immagine dello shower ora oppure alla fine quando parli di ENC 
Over the last few decades, hybrid pixel detectors have been extensively used in High Energy Physics (HEP) experiments \cite{kolanoski2020particle,rossi2006pixel}. These detectors showed their  ability to cope with the harsh Large Hadron Collider (LHC) environment and to efficiently reconstruct collision events with a very high density of particles  \cite{rossi2006pixel}. 
The main advantage of hybrid pixel detectors is the possibility to use sensors of different materials (e.g. gallium arsenide, cadmium  telluride, diamond \cite{kolanoski2020particle}) or to separately optimize sensors  and electronics. These options have been extensively exploited in many HEP experiments and medical applications \cite{delpierre2014history}. 

More recently, monolithic architectures, in which the sensor and the electronics are fabricated on the same silicon wafer, have  been proposed \cite{snoeys2013monolithic,  ttpetdemo, sigedesign2019, sige2019, sige2020}. Even if many design challenges made these architectures not as widely exploited in HEP applications as  hybrid ones \cite{snoeys2013monolithic}, they represent an efficient solution to reduce the production costs and to  minimize the material budget \cite{valerio2013electronic}, thus paving the way to instrument experiments with higher precision detectors at an affordable cost (see for example \cite{aliceITS}). 

This paper presents a monolithic pixel detector test-chip developed to study different design solutions to be used for the high-precision pre-shower upgrade of the ForwArd Search ExpeRiment (FASER) \cite{FASER} at CERN. 
The  FASER experiment will search for the production of low-mass long lived particles (LLPs) not foreseen by the Standard Model of particle physics, such as dark photons and axion-like particles (ALPs)~\cite{feng2018forward,FASER,feng2018dark,feng2018axionlike}. 
The detector is installed in a service tunnel 480 m downstream from the ATLAS experiment
to make use of
the huge pion flux that collisions produce in the direction of the LHC beams.
The exploitation of the decay of this very large number of  pions, that is not accessible to the other experiments,
will therefore allow the extension of the LHC physics programme. 
Although the  FASER detector is very well-suited for LLPs that decay in charged particles ($e^-e^+$ pairs in the case of dark-photon decays), at present it would not be able to discriminate the two photons  produced by the decays of ALPs. For this reason, the FASER Collaboration  decided to upgrade
the detector with a high-precision tungsten-silicon pre-shower that will be able to identify the electromagnetic showers produced by the two photons from ALP decays  with energies of up to few TeV at a distance of 200 $\upmu$m. The ultimate goal is to drastically reduce  the backgrounds to this process.

%an enhancement of pre-shower detector is in progress: the proposed high-granularity pixel detector for the pre-shower structure will be characterized by a time resolution in the order of 100 ps and a density of $\sim$10k pixels/cm\textsuperscript{2}. The timing information will be used to indentify large deposition centers produced by low energy electrons and back-splash from the calorimeter. 
The 130 nm SiGe BiCMOS technology  by IHP Microelectronics,
used at the University of Geneva to develop monolithic silicon pixel sensors for timing purposes~\cite{sigedesign2019,ttpetdemo,sige2019,sige2020},
was chosen to produce the ASIC for the  FASER high-granularity pre-shower. 
%FASER can be considered as one of the first projects in which monolithic pixel detectors designed in a SiGe BiCMOS technology are exploited in a HEP application. 
The prototype chip presented here was designed to study the integration of the front-end electronics (or part of it) in the sensor area, a solution explored to minimise the detector inactive area, that is a potential limiting factor of monolithic architectures.

The paper is organized as follows: Section \ref{sec:test_chip} provides a detailed description of the test chip; Section \ref{sec:architecture} focuses on the analog front-end system and its impact on the detector performance, as well as on the  architecture variants introduced in the test chip to choose the best to be implemented in the final pre-shower ASIC; Section \ref{sec:measurements} shows the results of the prototype-chip measurements. All the simulation results shown in the paper have been produced with Cadence Spectre.

\section{The prototype chip}
\label{sec:test_chip}

\begin{figure}[!t]
	\centering
	\includegraphics[width=5in]{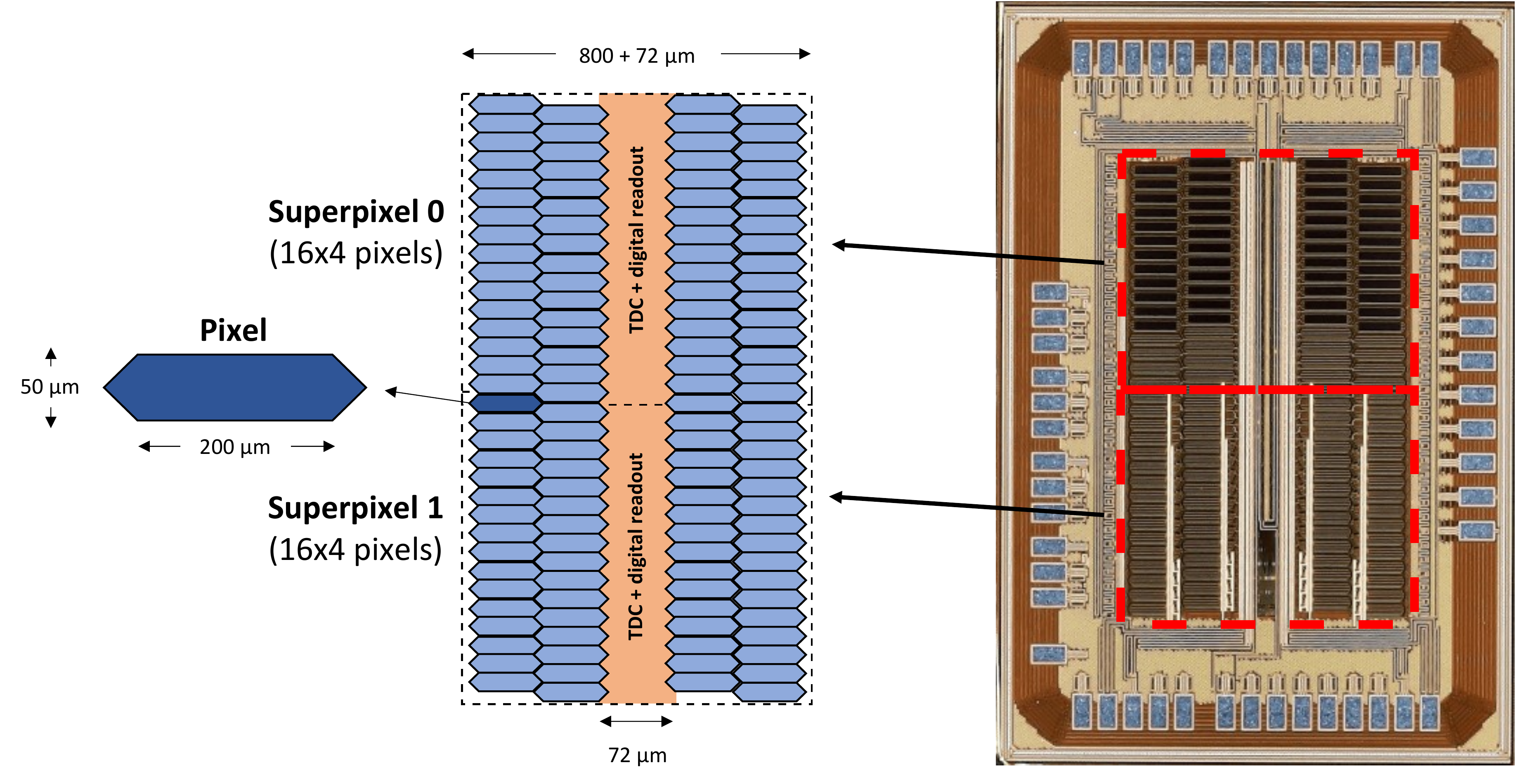}%
	\caption{Super-pixels and pixel size (left) and a photograph of the prototype ASIC (right); the ASIC total area is  1.7$\times$2.6 mm$^2$.}
	\label{chip_superpixel_faser_photo}
\end{figure}

The pixel-detector prototype described in this work is characterized by a monolithic structure, hence the pixelated sensitive area is integrated in the same chip with the front-end electronics. 
The purpose of this prototype is to study the integration of different front-end configurations in the pixel area and validate possible designs that can be used for the pre-production ASIC of the FASER pre-shower. 
Several front-end variants were designed to characterize the routing distribution and to understand how to minimize self-induced noise and cross-talk. 
Gain, noise and stability  of front-end variants were analysed in order to choose the best configurations. The target performance is 1 fC input charge discrimination threshold and 150 mW/cm\textsuperscript{2} analog power consumption.

A photograph of the prototype chip  is shown in Figure \ref{chip_superpixel_faser_photo} (right).
The pixel matrix of the ASIC is composed by two superpixels, as seen in Figure \ref{chip_superpixel_faser_photo} (left), each featuring 16 $\times$ 4 pixels with a 200 $\times$ 50 $\upmu$m$^2$ area\footnote{The pixel area and shape implemented in this prototype were those considered for the FASER high-precision pre-shower at the time of the submission of this ASIC. Successive full-simulation studies have shown that the optimal pixel size is hexagonal with 65 $\upmu$m side (corresponding approximately to a pixel pitch of 100 $\upmu$m), which has been adopted for the final ASIC.}, a Time-to-Digital Converter (TDC) and a digital readout logic placed in the 72 $\upmu$m thick region highlighted in orange in Figure \ref{chip_superpixel_faser_photo}.  
\begin{figure}[!t]
	\centering
	\includegraphics[width=2.5in]{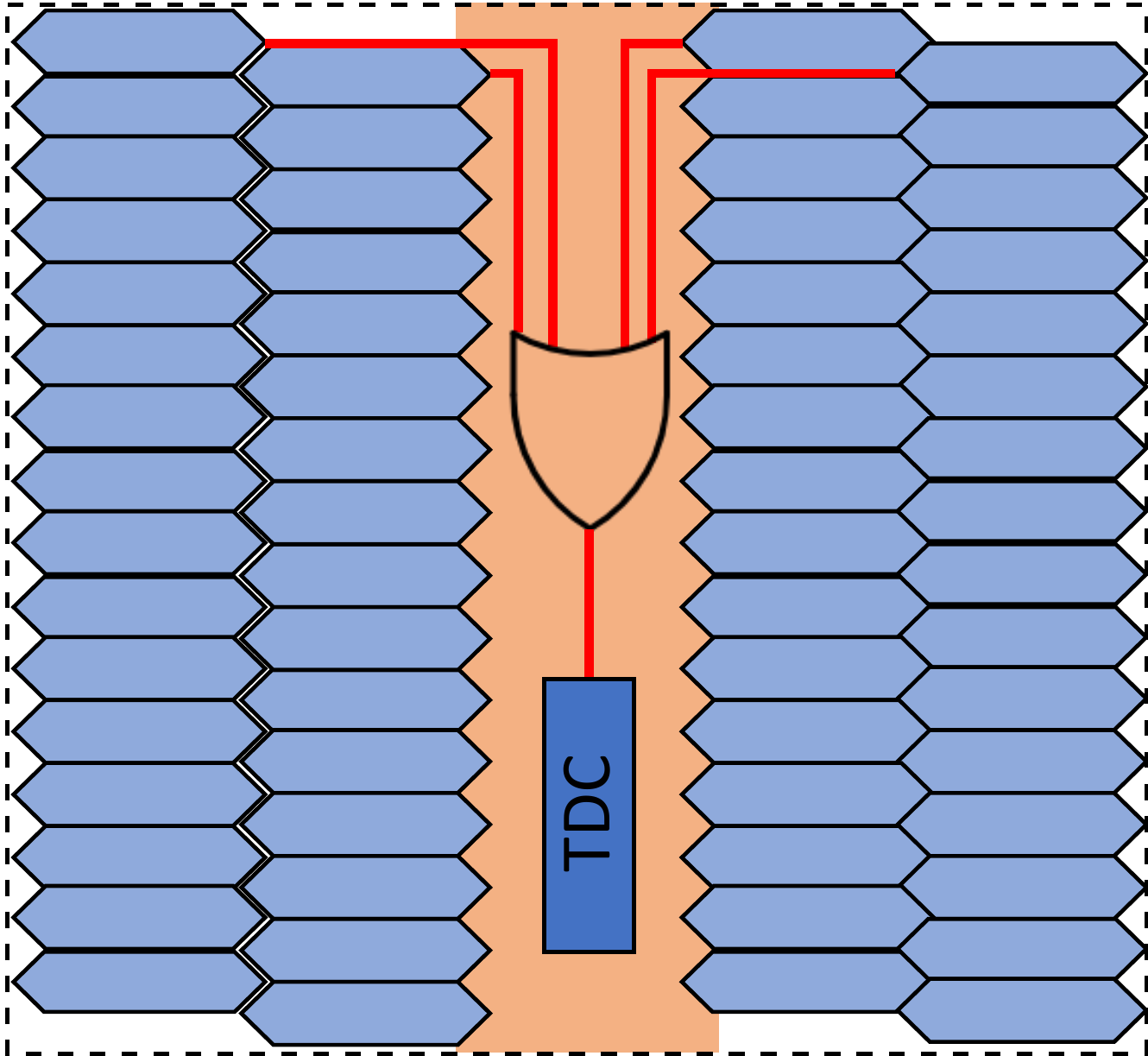}
	\caption{Drawing of a superpixel of 4$\times$16 pixels. The four pixels in a row are multiplexed in one channel of the TDC.}
	\label{multiplexing}
\end{figure}
The active area of the pixels, i.e. the region in which the particles are sensed, has the shape of an elongated hexagon. The reason behind the choice of this shape  is to have 120$^\circ$ instead of 90$^\circ$ angles at the edges of the pixel sensitive areas  to reduce the electric field in these zones and thus the risk of an early breakdown in the pixel matrix \cite{ueda1985breakdown,sze1966effect,sige2019}. In this way, it is possible to bias the pixels with higher voltages potentially leading to better  performance \cite{paolozzi2015development}.
%si riferisce alle prestazioni di timing con bias piu alti e campi uniformi

The TDC features 16 channels, one per each superpixel row. Hence, the output of pixels in the same row are multiplexed as in Figure \ref{multiplexing} making each superpixel able to distinguish simultaneous events only in the vertical direction. This architectural choice enables integrating a smaller number of TDC channels in the chip with a consequent reduction of the power consumption and inactive area within the superpixel, that in the case of this prototype chip is less than 9 $\%$ of the total superpixel surface. Pixel multiplexing is shown in Figure \ref{multiplexing}. If more than one signal occurs in a row of a super-pixel  within a time window shorter than the minimum time the system needs to perform two consecutive conversions, then the readout logic will store the positions of all firing pixels. However, only the timing information relative  to the first firing pixel will be stored.
%However, the readout logic has been designed to store also the positions of the other firing pixels within the same time window.
%The architecture of the front-end system variants and their characteristics are described in the Section \ref{sec:architecture}.

%~~[\textcolor{red}{manca la figura compression.pdf  e quel testo...}]

\iffalse
\\The superpixels of the final chip will also be characterized by a data compression logic (the block diagram is depicted in Figure \ref{compression}). During the readout, each row of the superpixel can produce approximately 40 bit of data including timing information (associated to the particles sensed by the pixels). In order to significantly reduce the amount of data being read out, a flag register (F in Figure \ref{compression}), indicating whether a pixel has been hit, can be used to skip all the rows with no events. 
\fi
%\begin{figure}[!t]
%	\centering
%	\includegraphics[width=3.5in]{figures/tdc.pdf}
%	\caption{Block diagram of the TDC designed for the demonstrator chip.}
%	\label{tdc}
%\end{figure}

\iffalse
\begin{figure}[!t]
	\centering
	\includegraphics[width=3.5in]{figures/compression.pdf}
	\caption{Block diagram of the superpixel data compression logic.}
	\label{compression}
\end{figure}
\fi
\section{Front-end architecture}
\label{sec:architecture}

\subsection{Pre-amplifier and design choices}
\label{sec:pre-amp}
\begin{figure}[!t]
	\centering
	\includegraphics[width=5.5in]{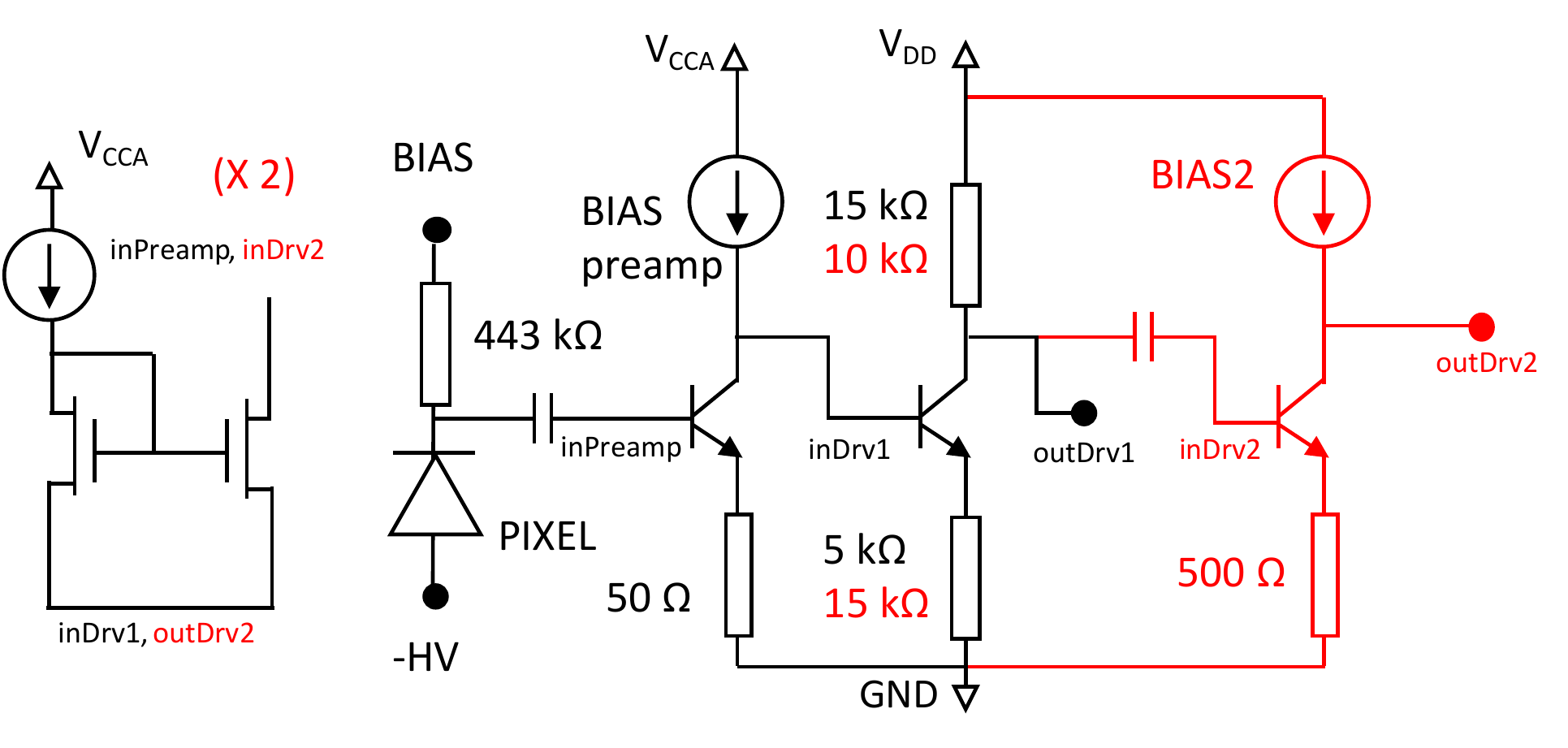}
	\caption{Pre-amplifier architecture. The stage and resistance values highlighted in red refer to the inverting solution that characterizes only one of the flavours adopted for the test chip.}
	\label{front_end}
\end{figure}
The sensor implemented in this prototype chip is
a PN junction operating in reverse bias. In this mode, the anode is connected to the p-doped substrate of 50 $\upOmega$cm bulk resistivity while the cathode is an n-well with an elongated hexagonal shape, as described in Section \ref{sec:test_chip}.  When  a negative voltage  HV=-120 V is applied, a depletion region of approximately 20-25 $\upmu$m is created in the substrate. Such depletion region leads to the generation of 1200-1800 electron-hole pairs \cite{beringer2012particle} when a minimum-ionizing particle passes through the sensor.
 
The pre-amplifier is the first stage connected to the sensor (depicted  as a diode in reverse bias in Figure \ref{front_end}). 
The pixel n-well is biased at a low voltage through a 443 k$\upOmega$ resistor. 
The pre-amplifier features a single-ended BJT-based first stage with active load. The latter provides a bias current of few $\upmu$A and is implemented with a PMOS transistor. Since the connection with the pixel sensor is in AC, an additional bias is needed for the base current of the bipolar transistor. For this purpose the block on the left part of Figure \ref{front_end} is used, in which the MOSFET on the right, used as a feedback impedance, is able to show a resistance of several hundred k$\upOmega$ (simulations highlight a value of approximately 670 k$\upOmega$ for a feedback bias current of 30 nA). The AC coupling capacitor is implemented with several PMOS transistors whose body terminals are connected to the n-doped cathode of the pixel and the gates, shorted to sources and drains, to the base of the BJT. 

The SG13G2 130 nm SiGe BiCMOS technology by IHP gave the design team the possibility to exploit SiGe-based Heterojunction Bipolar Transistors (HBTs) for the design of the front-end system. Indeed, every BJT presented in the schematics of this paper feature heterojuctions. The importance and the role of HBTs for the performance of time resolved pixel detectors is highlighted in \cite{paolozzi2015development}.

The first stage of the front-end system behaves as a charge amplifier, capable of producing output voltage signals directly proportional to the charge injected at the input. This behaviour is justified by the bandwidth of the amplifier and the characteristics of input signals. The amplifier is characterized by a first order pole that can be calculated using the Miller theorem as follows
\begin{equation}
\label{eq:pole}
    f_p\approx\frac{1}{2\pi R_{in}(|A_v|+1)C_F},
\end{equation}
where $R_{in}$ is the input resistance of the pre-amplifier, $A_v$ is the voltage gain and $C_F$ is the feedback capacitance between the base and collector of the HBT (or drain and source of the feedback MOS of Figure \ref{front_end}). Simulations show that the main contribution of $C_F$ is given by the base-collector capacitance of the BJT and it is approximately $C_F\approx C_{BC}\approx$ 2fF. Moreover, considering an input resistance $R_{in}\lesssim$ 100 k$\upOmega$ and a voltage gain $A_v$ in the order of few tens, the pole described in Equation \ref{eq:pole} will lay in the 100 MHz range. Because of the above-mentioned 20-25 $\upmu$m depletion of the substrate, the input signals of the pre-amplifier will have rise times in the order of few hundreds of picoseconds. Therefore, the spectrum of these signals will be over the first pole of Equation \ref{eq:pole} and the amplifier will work in integration regime i.e., as a charge amplifier. The work proposed in \cite{paolozzi2015development} emphasizes the advantages of this approach for the timing performance of the front-end. Indeed, a smaller bandwidth is useful to reduce the noise and for a better suppression of the crosstalk related to noisy external sources (e.g., digital periphery).

For a charge amplifier, the Equivalent Noise Charge (ENC) represents a crucial parameter to minimize for the optimization of the timing performance \cite{gatti1986processing}. It is defined as the input charge to be injected in an ideal and noiseless version of the amplifier that produces an output characterized by the same root-mean-square of the output noise of the real amplifier. From this definition and considering that the Signal-to-Noise Ratio ($SNR$) can be obtained by $SNR=Q_{in}/ENC$, where $Q_{in}$ is the input charge of the amplifier, it is possible to express the jitter contribution of the electronics $\sigma_{elec}$ to the total time resolution as: 
\begin{equation}
\label{eq:sigma_elec}
    \sigma_{elec}=\frac{\sigma_v}{dV/dt}\approx \frac{t_{rise}}{SNR}=\frac{t_{rise} ENC}{Q_{in}},
\end{equation}
where $\sigma_v$ indicates the output noise of the circuit, $t_{rise}$ is the rise time of the output signal of the pre-amplifier and $dV/dt$ is its slope.
Moreover, \cite{gatti1986processing} demonstrated that amplifiers with bipolar transistors with high current gains show better performance in terms of ENC compared to MOS-based front-end for sub-nanosecond shaping time.

The second stage of the pre-amplifier is used to increase the voltage gain of the system by a factor $A_{v2}\approx R_C/R_E$ where $R_C$ and $R_E$ represent the collector and emitter resistances. It is also lowering the output impedance of the pre-amplifier, making it more robust to routing. The coupling with the first stage is in DC, hence this stage does not need another bias system for the base current.

The third stage, highlighted in red in Figure \ref{front_end}, is an additional gain stage used to produce an output signal (outDrv2) with the same polarity of the input of the pre-amplifier system (inPreamp). This particular configuration is featured only in one of the pre-amplifier variants that have been integrated in the test-chip. The design process of this configuration aimed to implement a structure in which the stability of the amplifier was the main optimization parameter to take into account. Indeed, outDrv2 follows the same behaviour of inPreamp inducing a negative feedback that is meant to avoid unwanted oscillations, making this architecture suitable for an integration inside the active area of the pixel. This configuration presents a second stage with a reduced gain compared to the other (10 and 15 k$\Omega$ instead of 15 and 5 k$\Omega$ as indicated by the red labels of the resistors in Figure \ref{front_end}) in order to further reduce the impact of the output (outDrv1) that shows a positive feedback coupling with the pixel. 
The other pre-amplifier flavours only feature the part of the circuit depicted in black in Figure \ref{front_end}. The differences among these configurations will be described in Section \ref{sec:flavours}.
More details about the techniques to improve the stability of the system are reported in Section \ref{sec:layout}.
\begin{figure}[!t]
	\centering
	\includegraphics[width=4in]{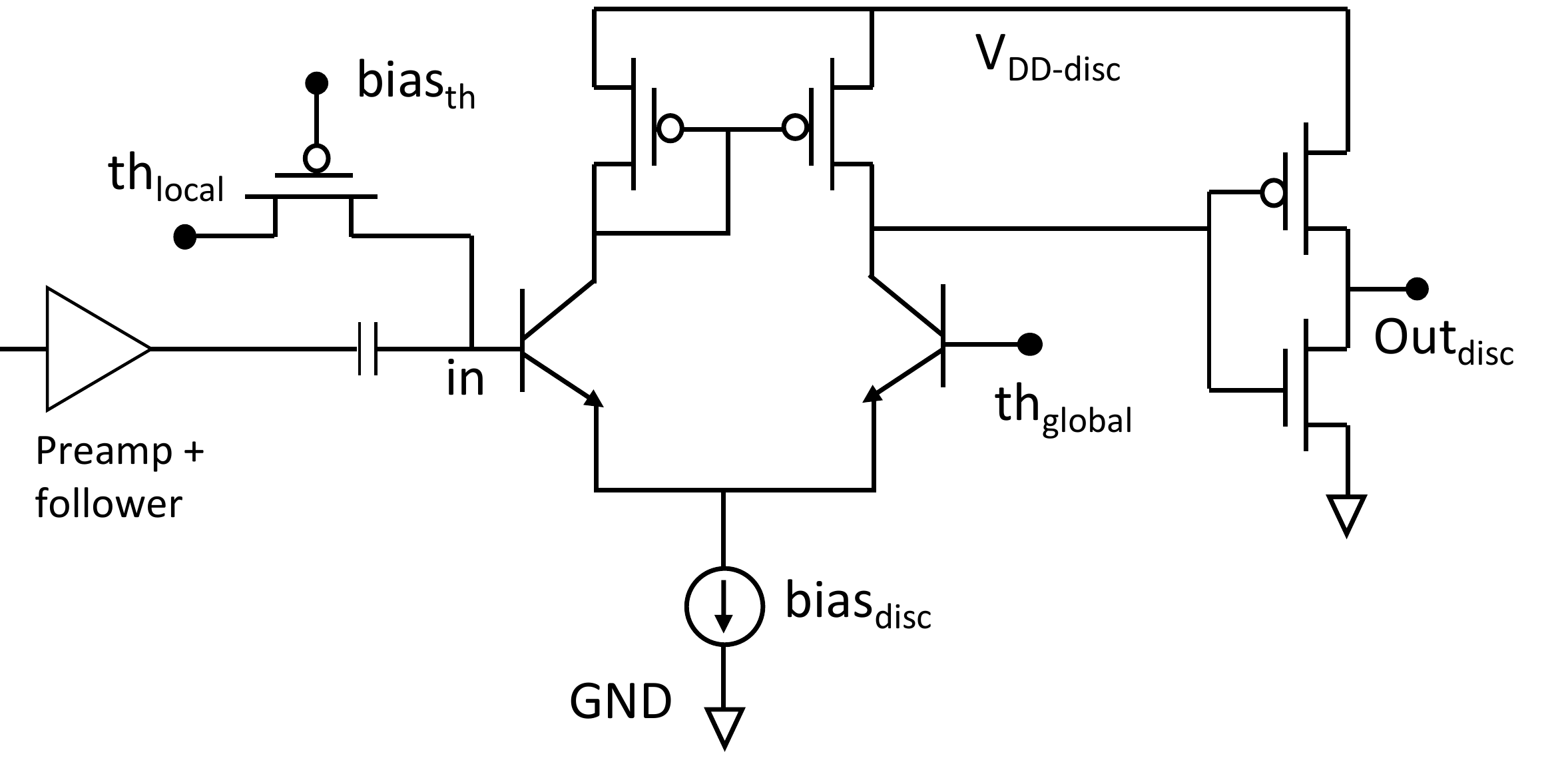}
	\caption{Discriminator schematic and connection to the pre-amplifier.}
	\label{disc}
\end{figure}

%explanation of the architecture
%motivation behind the SiGe technology (see thesis also and Lorenzo)
%OPTIONAL: small-signal analysis circuit and expected performance (in agreement with the measurement)

\subsection{Discriminator}
\label{sec:peak_discriminator}
The schematic of the discriminator chosen for the front-end is reported in Figure \ref{disc}. The differential pair is designed  also in this case using SiGe HBTs with a PMOS-based active load. The threshold of the circuit is set through the global threshold signal $th_{global}$, distributed to the front-end of every pixel in the chip, and a local threshold $th_{local}$. As it will be explained in Section \ref{sec:layout}, the design of the discriminator plays an important role in the stability of the front-end, especially when the system is integrated in pixel. 
%motivations to be connected to the end of the beginning of the section 
%analysis and simulation 
%brief overview of the flavours adopted in the test-chip 
%(obviously some stuff can differ and the from what expected in the measurements, like the one related to the architecture with more buffer to make the 
%structure stable)
\begin{figure}[!t]
	\centering
	\includegraphics[width=5in]{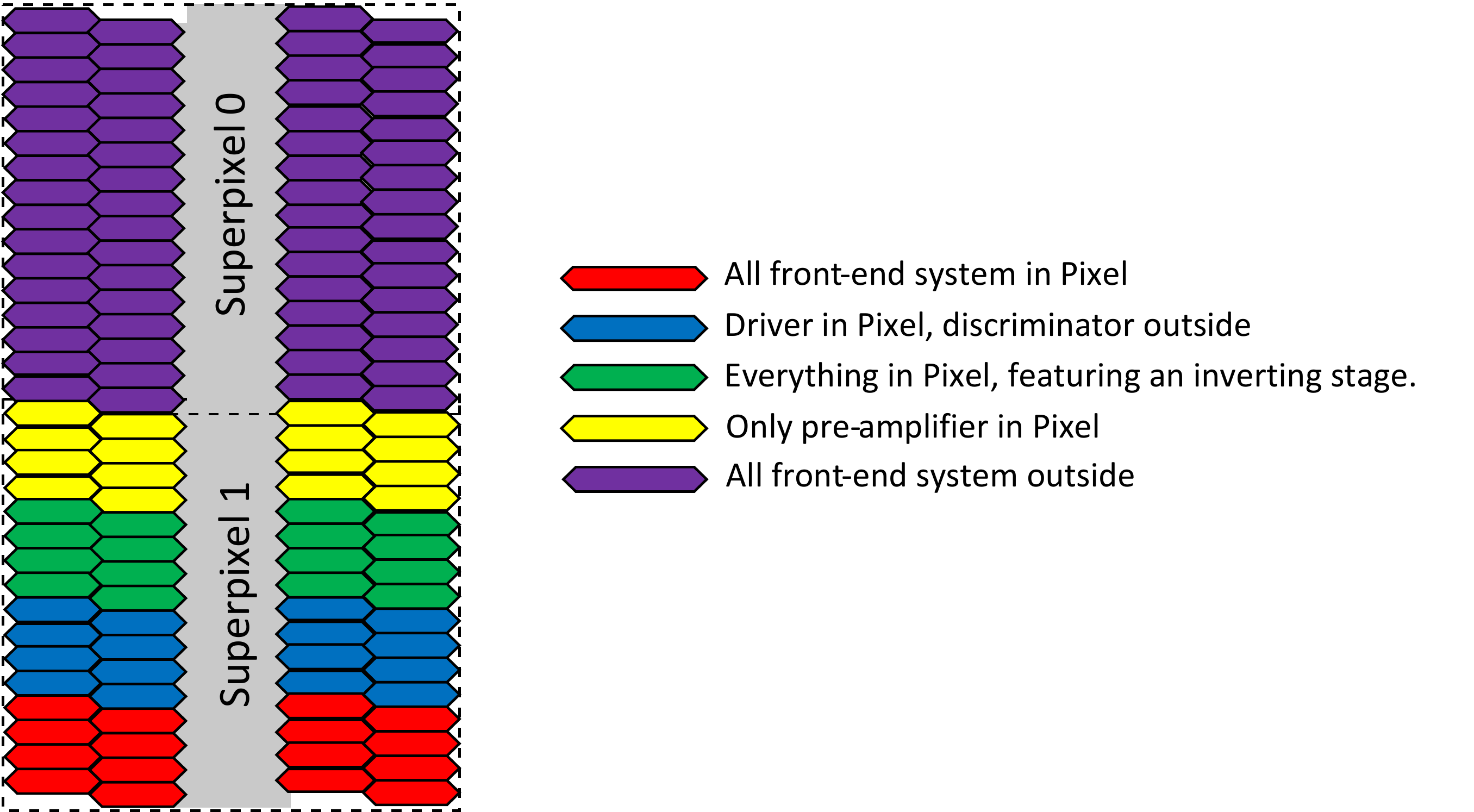}
	\caption{Distribution in the pixel matrix of the front-end variants of this test chip.}
	\label{flavours}
\end{figure}
%all the flavors are in Run_2020_03_FASER_prototype- testbenches

\subsection{Flavours adopted for this prototype chip}
\label{sec:flavours}
The  aim of this prototype chip was to perform an analysis of several front-end configurations and to choose the best solution for the final full-reticle ASIC of the FASER pre-shower. These configurations are characterized by different degrees of integration in pixel of the electronics. In general, integrating circuits inside the sensitive area can lead to an increase of the detector capacitance and, consequently, to a reduction of the SNR. Therefore, the noise performance of the versions of the front-end with many blocks in pixel were expected to show worse performance in terms of noise. Further details will be provided in Section \ref{sec:comparison}.

Figure \ref{flavours} shows the floorplan of the front-end variants:
\begin{itemize}
    \item the first one, reported in purple, features the whole front-end outside the pixel. Because of the demanding specification of the final FASER ASIC in terms of dead area and pixel density, this version will not be inserted in future iterations of the chip. However, it was integrated in the presented prototype ASIC to compare its performance with the other versions of the front-end: this solution was expected to be the least critical one for the stability and noise thus it can be used as a reference for the evaluation of other configurations.
    \item the pixel reported in yellow feature a front-end with only the pre-amplifier inside the pixel. The gain of this architecture was expected to be smaller than others since increasing the length of the routing that connects the output of the pre-amplifier with the driver can significantly reduce the band-width and the gain of the system.
    \item in blue and green the pixel configurations with pre-amplifier and driver integrated in the sensitive area are reported. In particular, the green ones indicate the solution with an additional inverting stage to further improve the stability (reported in red in Figure \ref{front_end}). 
    \item finally, the configuration reported in red, has all its blocks, including the discriminator, integrated in the pixel well. As it will be highlighted in Section \ref{sec:layout}, particular care was dedicated to the layout of this architecture in order to reduce self-induced oscillations. 
\end{itemize}
During the design process, the last three solution were expected to be the best candidates to be integrated in the final chip of the FASER detector because they are the least demanding in terms of dead area. Their performance and a more detailed comparison of the configurations are reported in Section \ref{sec:comparison}.

\iffalse
The first four from the bottom are organized in groups of 16 pixels (4 rows). 
The first one, reported in red, has all its blocks, including the discriminator,  integrated in the pixel well. As it will be highlighted in Section \ref{sec:layout}, particular care was dedicated to the layout of this architecture in order to reduce self-induced oscillations. The other architectures feature a progressive extraction of blocks from the pixel area: the discriminator outside the pixel (blue); the additional inverting stage shown in Figure \ref{front_end} outside the pixel (green); only the pre-amplifier inside the pixel (yellow);  the whole front-end outside the pixel (purple). 
\fi
%Finally, the remaining 48 pixels were connected to front-end circuits outside the pixels and used as analog channels.  
%sensor - active area - pixel - electronic blocks

\subsection{Cross-talk compensation and layout}
\label{sec:layout}

\begin{figure}[!t]
	\centering
	\includegraphics[width=4.1in]{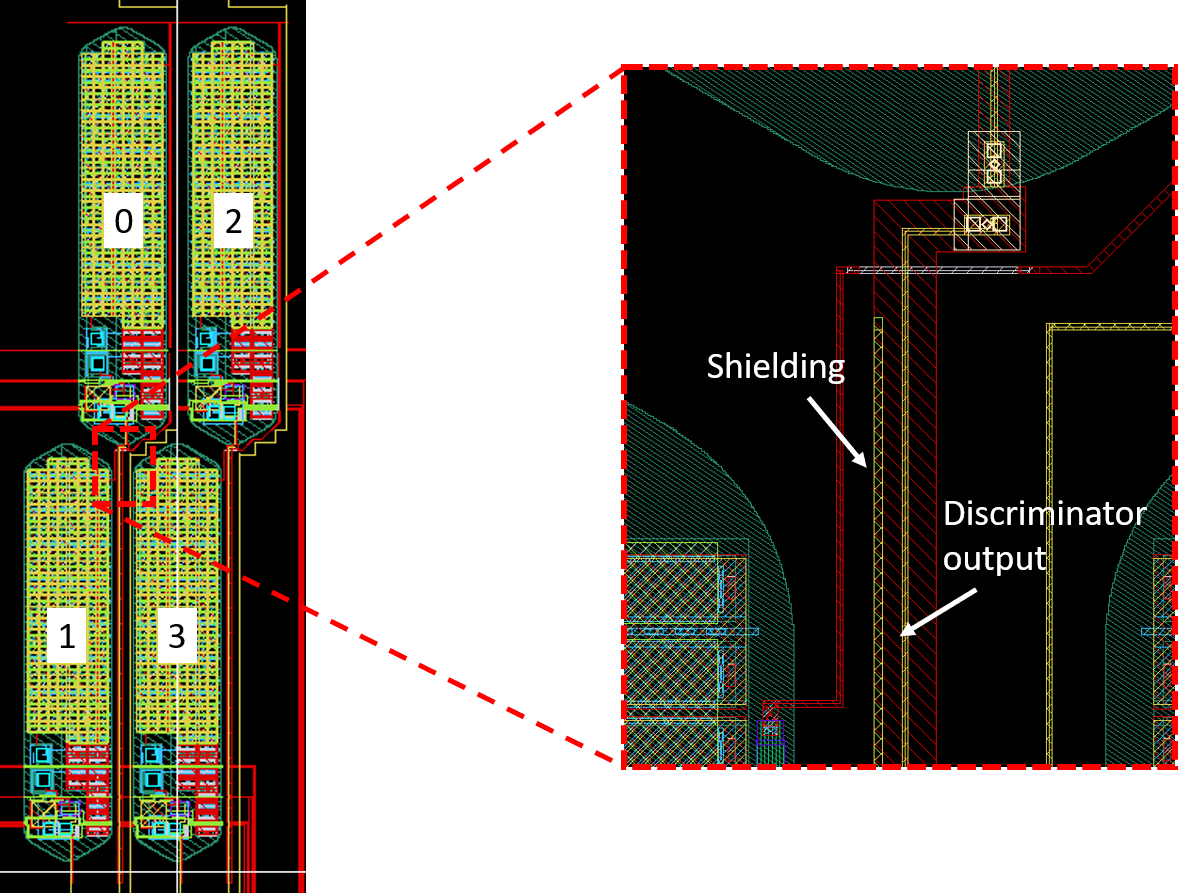}
	\caption{Layout of 2$\times$2 pixels (left) and zoom (right) on the shielding line to reduce the cross-talk between the output of the discriminator and the adjacent pixel wells.}
	\label{cross_talk_layout}
\end{figure}
\begin{figure}[!t]
	\centering
	\includegraphics[width=6in]{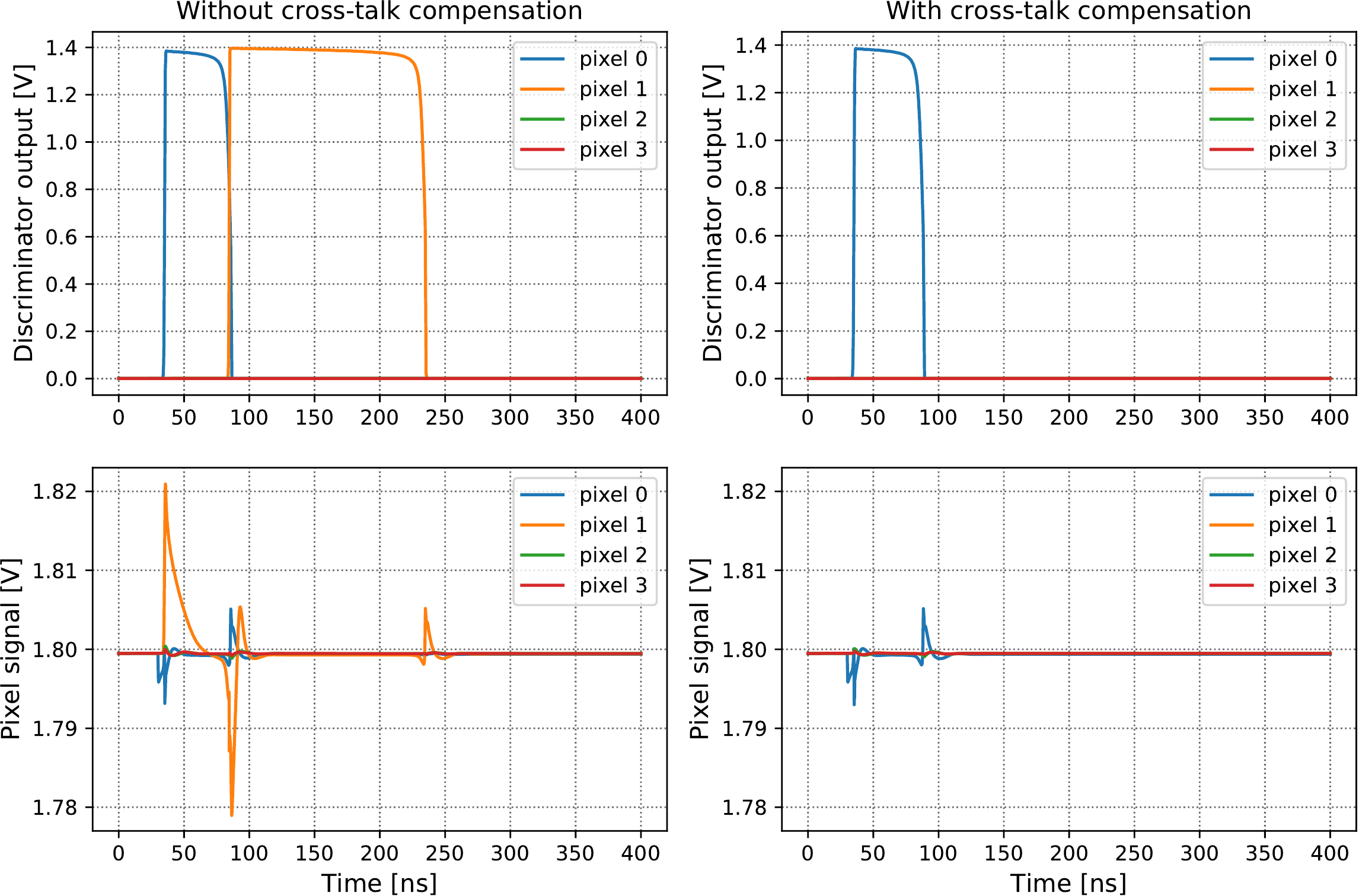}
	\caption{Discriminator output and pixels signal with (right) and without (left) cross-talk protection lines for an input charge of 0.5 fC.}
	\label{cross_talk}
\end{figure}
\begin{figure}[!t]
	\centering
	\includegraphics[width=5in]{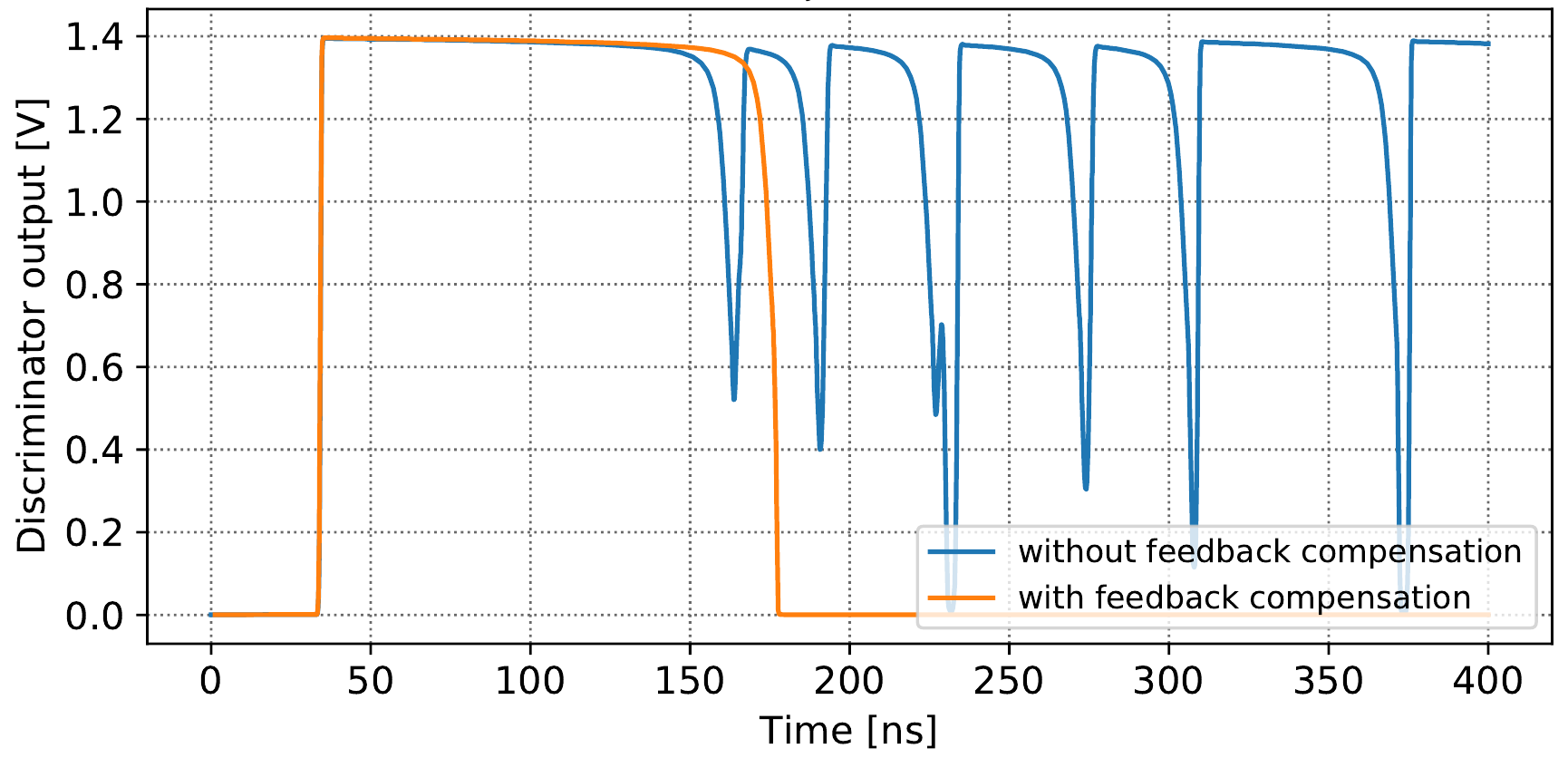}
	\caption{Discriminators output behaviour with (orange) and without (blue) self-induced noise compensation lines.}
	\label{self_compesation}
\end{figure}
\begin{figure}[!t]
	\centering
	\includegraphics[width=2.1in]{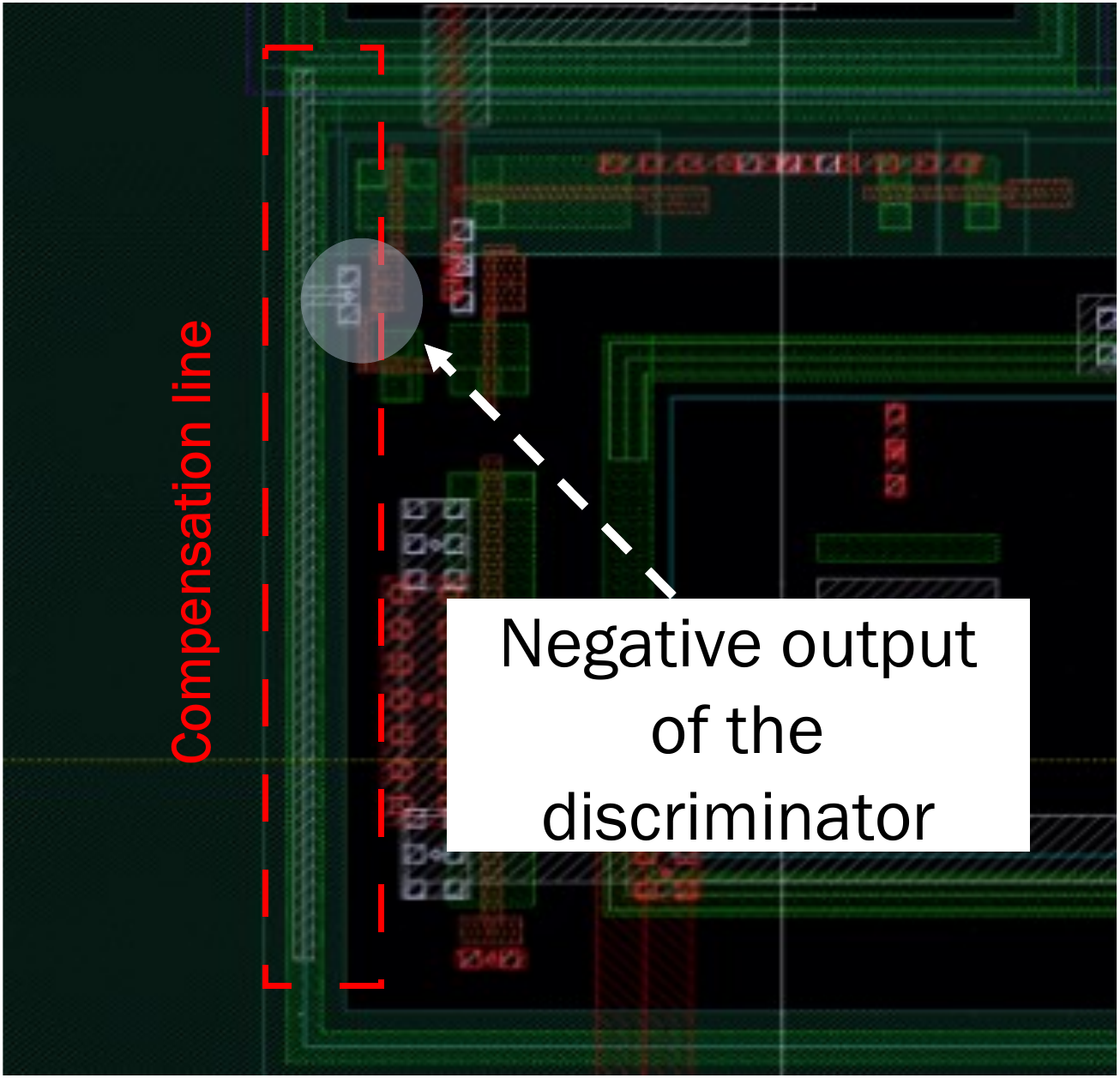}
	\caption{Example of self-induced noise compensation line. The metal2 line is connected to the negative discriminator output to increase the coupling with the pixel well and avoid self oscillation induced by the effect of the positive output.}
	\label{self_compensation_layout}
\end{figure}
A crucial part of the design process  focused on the optimisation of the layout. An extensive simulation campaign was launched to analyse the performance of the chip. Layouts with 2$\times$2 pixels sub-matrices featuring various front-end configurations were simulated to evaluate the effects of  coupling between  discriminator output lines and  neighbouring pixels, as seen in Figure \ref{cross_talk_layout} (left). The analysis showed that, if no shielding line is included in the layout, the falling edge of the discriminator output of pixel 0 of Figure \ref{cross_talk_layout} can induce a spurious hit on another pixel (in this case,  pixel 1) as it is reported in the plots of Figure \ref{cross_talk} (left). The shielding lines avoid this problem, as seen in Figure \ref{cross_talk} (right), without any significant impact on the discriminator performance. %The efficiency of the shielding lines were confirmed by the measurement results.

As mentioned in Section \ref{sec:flavours}, the front-end configuration that present a complete integration of the discriminator inside the pixel active area may be critical for the stability of the system. Indeed, as clearly shown in Figure \ref{self_compesation}, the coupling between the discriminator output and the pixel could generate a positive feedback leading to unwanted oscillations of the front-end chain. In the present prototype the problem  was solved  inverting the polarity of the output of the discriminator and increasing its coupling with the pixel exploiting additional metal lines as shown in Figure \ref{self_compensation_layout}. This compensation technique led to a correct behaviour of the front-end  (orange curves in Figure \ref{self_compesation}).

\section{Measurements}
\label{sec:measurements}

\begin{figure}[!t]
	\centering
	\includegraphics[width=5in]{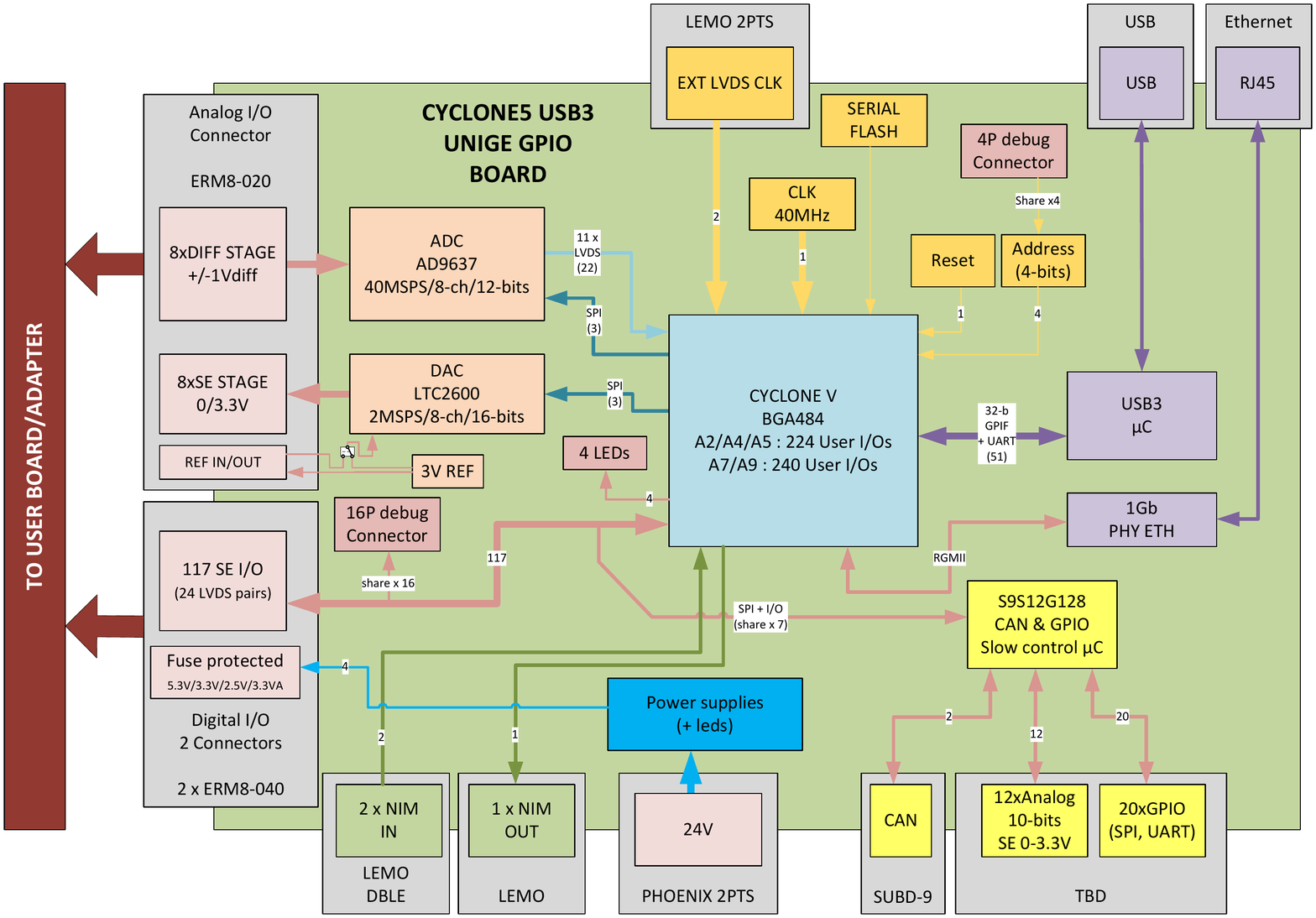}
	\caption{Schematics of the UNIGE USB3 GPIO readout board.}
	\label{gpio}
\end{figure}

%In this section, the results of the measurements performed on the test-chip will be described. 
The ASIC has been tested and qualified with the UNIGE USB3 GPIO system, depicted in Figure \ref{gpio}, that was initially developed by the engineering team of the D\'epartement de Physique Nucl\'eaire et Corpusculaire (DPNC) of the University of Geneva for the Baby-MIND experiment at CERN \cite{noah2016readout}. The GPIO system includes a readout board that uses an Altera Cyclon V FPGA that can  control several detectors at once. The system features a modular software framework customizable with JSON files that allow implementing GUIs for the control and readout of the ASICs under test.
\\All the measurements reported in this section were performed setting the power consumption of the chip at 144 mW/cm\textsuperscript{2} thus within the  150 mW/cm\textsuperscript{2} specification of the FASER experiment. 

\subsection{Calibration}

As shown in Figure \ref{disc}, the front-ends of this prototype feature a global threshold $th_{global}$ signal connected to the base of the negative input of the discriminators. In addition to this global threshold, each pixel also features a 6-bit Digital-to-Analog Converter (DAC) used to set the local threshold $th_{local}$. The role of this signal is to bias the positive input of the discriminator and move the threshold of the discriminators with respect to $th_{global}$. The $th_{local}$ of each front-end can be set independently of each other to compensate for mismatch effects and  guarantee that the equivalent threshold is as uniform as possible across the matrix.

\begin{figure}[!b]
	\centering
	\includegraphics[width=3.0in]{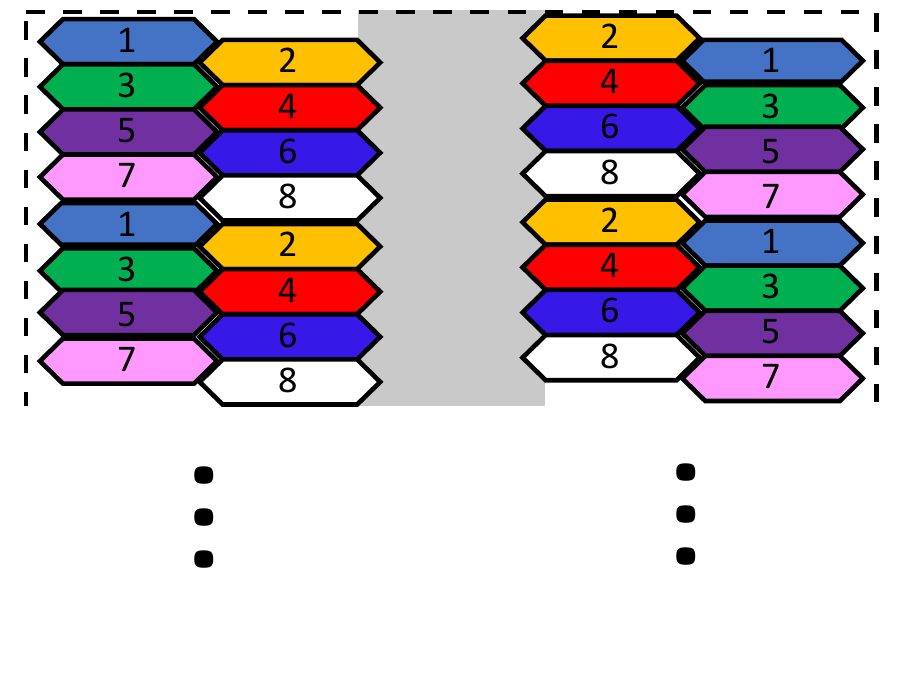}
	\caption{Mapping of the pixels in the eight groups used for the threshold calibration.
	%%\textcolor{red}{Fulvio, questa figura va migliorata e fatta con più precisione}
	}
	\label{cal_groups}
\end{figure}

For a certain value of $th_{global}$, the calibration algorithm is based on a scan of the local threshold for each pixel performed by changing the input of the associated DAC. In this way, it is possible to obtain the values of the input codes of the converter that make the discriminator output switch. Indeed, in a certain range of $th_{local}$ the baseline of the input will be so close to its threshold that the noise will activate the discriminator multiple times. The correct value of DAC input for the calibration is chosen such that the noise hit rate is low (in the 0.1 - 0.01 Hz range). The fastest way to perform this operation is to make a simultaneous scan of all pixels. However, the activation of several discriminators at the same time will produce peaks of absorption that can compromise the accuracy of the calibration process. Measurements highlighted that the difference between the threshold value obtained with the calibration of a single pixel and the one given by a simultaneous scan of the whole chip can be up to  22\% of the DAC dynamic range. For this reason, an alternative process 
was developed in which the pixels were divided in eight groups such that the calibration was performed independently for each group.
An efficient distribution of the mapping of these eight groups of pixels is displayed in Figure \ref{cal_groups}: the distance maintained among the pixels of the same group drastically reduces the calibration error to values up to 1 Least Significant Bit (LSB) of the local DACs, i.e. less then 3\% of their dynamic range.
%commento su plot e su quanto sia necessario mettere i local DAC
\begin{figure}[!t]
	\centering
	\subfloat[]{\includegraphics[width=2.90in]{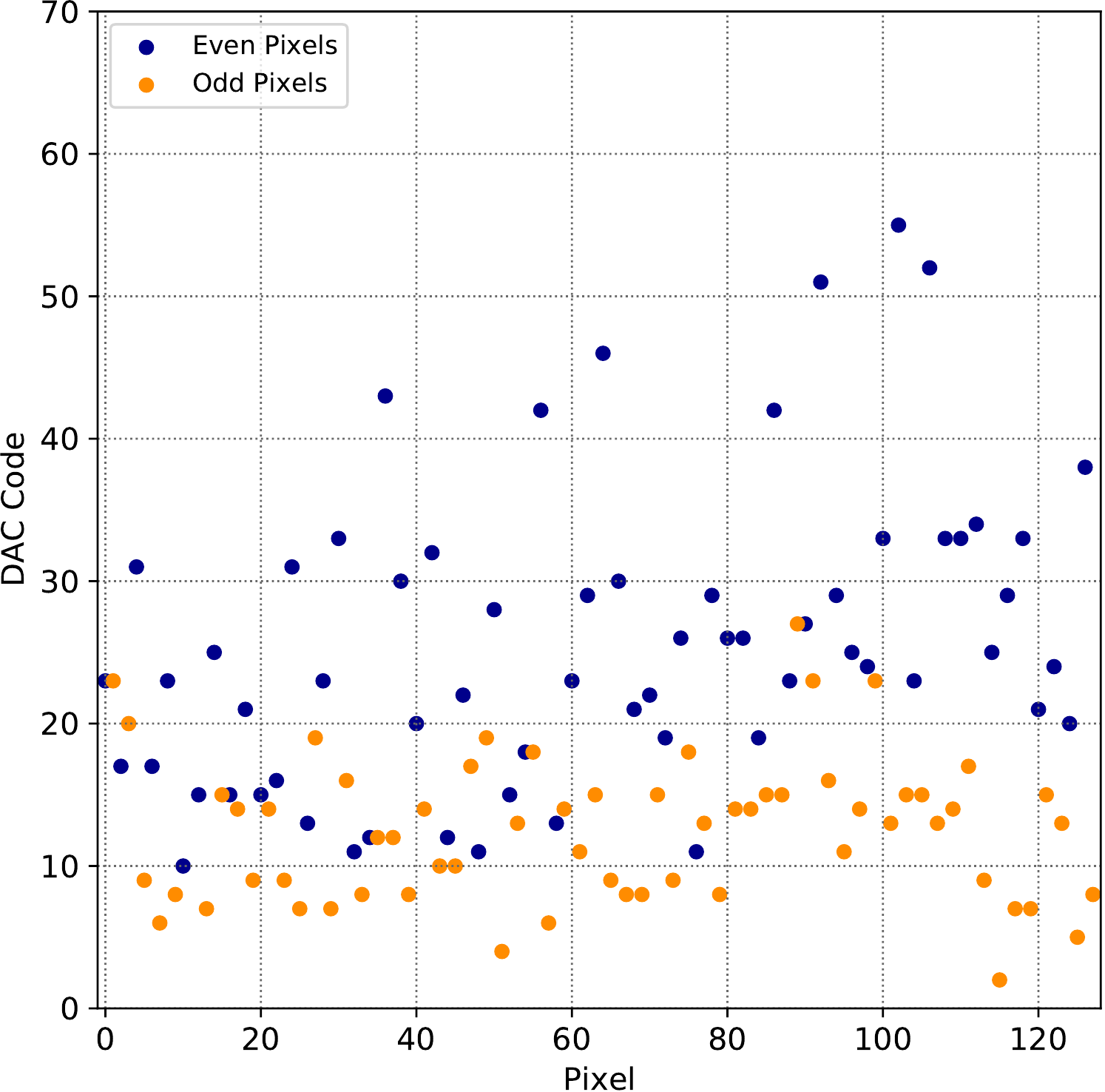}%
		\label{DAC_disp}}
	\hfil
	\subfloat[]{\includegraphics[width=2.97in]{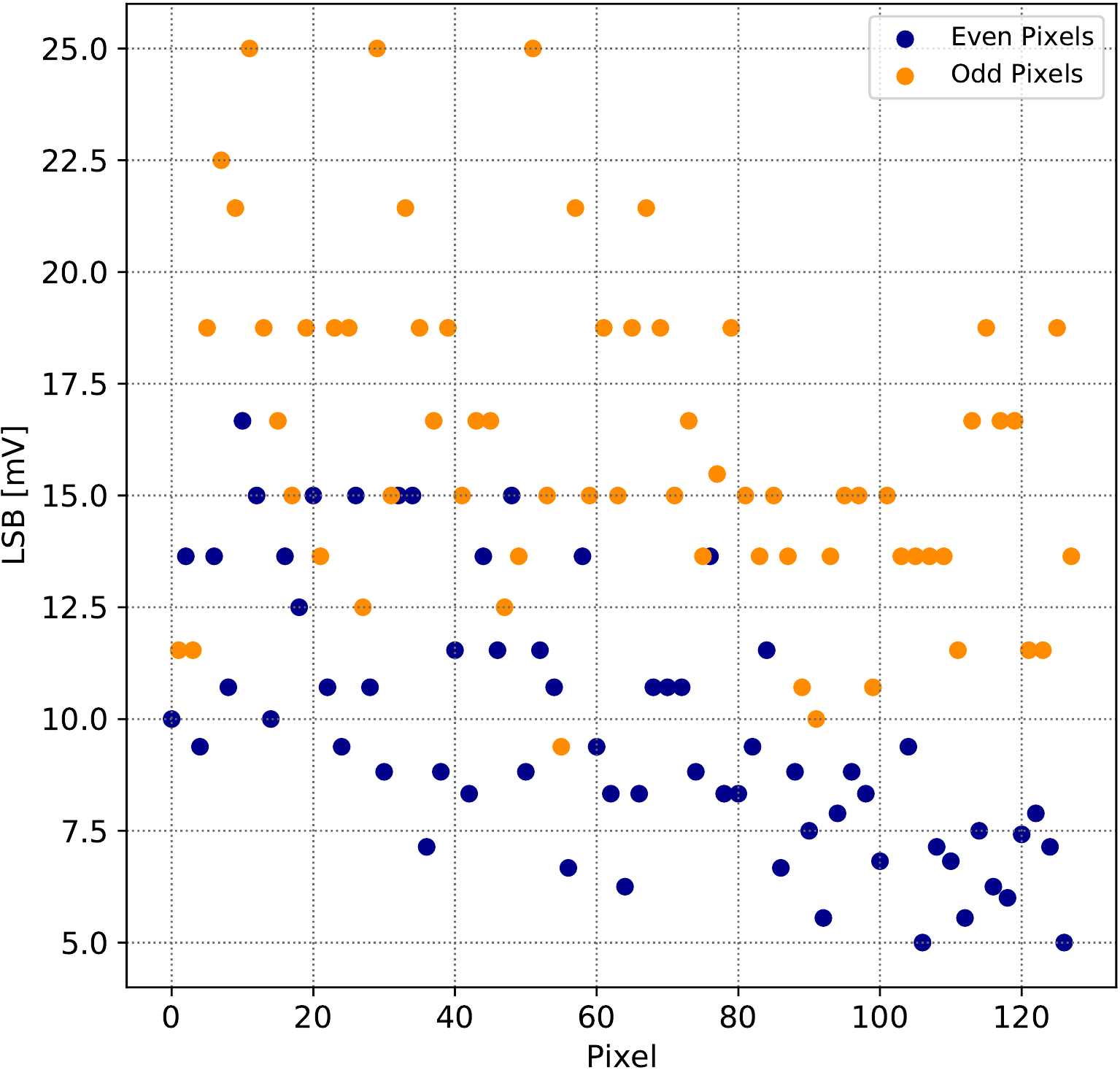}
		\label{LSB_disp}}
	\hfil
	\subfloat[]{\includegraphics[width=4in]{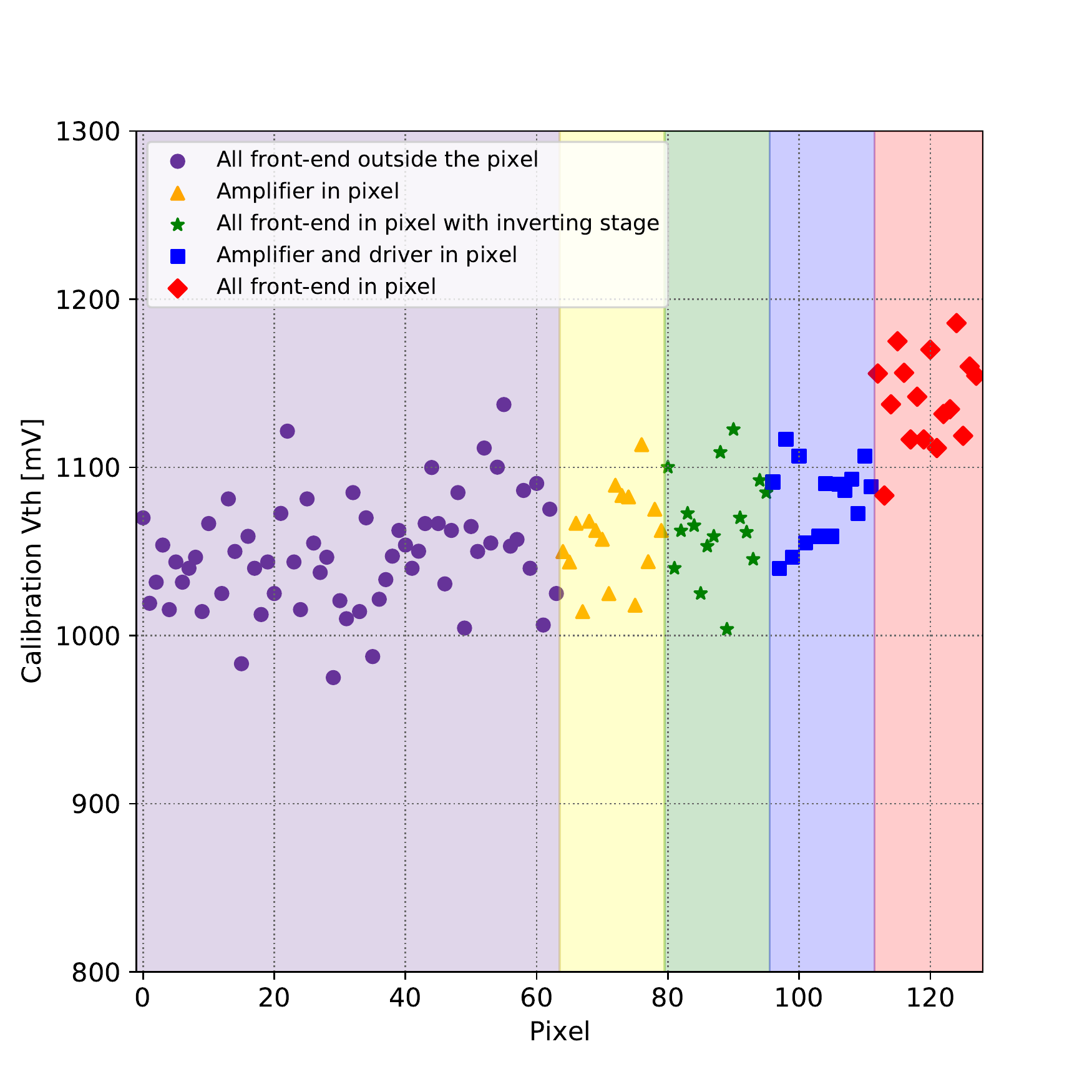}
	    \label{TH_disp}}
	\caption{DAC code (a), DAC average LSBs (b) and front-end threshold (c) distribution in the prototype chip. The x-axis indicates the pixel number. The pixel are labeled from bottom to top of Figure \ref{flavours} and split in even and odd on the right and the left part of the chip. The results in (a) were obtained  setting the global threshold at 1.1 V.
	}
	\label{dispertion}
\end{figure}

Figure \ref{dispertion} displays the results of the threshold calibration.  
Figure \ref{DAC_disp} shows the values of the DAC codes associated to the local threshold of the pixels while Figure \ref{LSB_disp} displays the distribution of the average LSB of the converters in the matrix. The LSB was calculated performing a calibration with four different values of the global threshold (0.95, 1.00, 1.05 and 1.10 V) and evaluating the corresponding calibration code for each pixel. At this point, the average LSB of the $i$-th pixel $LSB_i$ can be obtained as 
\begin{equation}
    LSB_i=\frac{\Delta(th_{global})}{\mu(\Delta(O_i))},
\end{equation}
where $\Delta(th_{global})$=50 mV is the global threshold step implemented for the measurement and $\mu(\Delta(O_i))$ is the average value of the difference between the code obtained for a given global threshold and the previous one. In these two plots, it is possible to highlight an asymmetry between the DAC outputs of the pixels on the left (even pixels) and right (odd pixels) portion of the matrix: this effect can be attributed to a gradient in the fabrication process of the chip and to asymmetries in the converters. The decreasing trend of the average LSB in Figure \ref{LSB_disp} can also be associated to process gradients and to voltage drops on the supply.

Figure \ref{TH_disp} shows the distribution of the equivalent threshold of all the front-ends in the chip. The threshold $V_{th,i}$ of the $i$-th pixel is calculated combining the information of the two previous plots as 
\begin{equation}
    V_{th,i}=V_{DD}-(10+O_i)LSB_i,
\end{equation}
where the factor 10 is associated to a current offset of 10 $LSB_i$ on the output of the converter. The asymmetry between even and odd pixels and the decreasing trend of the LSBs cannot be deduced by the plot of Figure \ref{TH_disp} because the latter only describes the dispersion of the equivalent threshold given by mismatches of the electronics, i.e. without the contribution of DAC.  The plot also highlights a threshold dispersion $\sim$30 mV (exact values are  reported in Table \ref{tab:sigma}). 
\begin{table}[!b]
    \centering
    \begin{tabular}{lc}
        \hline 
        Configuration & $\sigma_{V_{th}}$ [mV]\tabularnewline
        \hline 
        All f.e. outside pixel & 32.3 \tabularnewline
        Only pre-amp. in pixel & 26.9  \tabularnewline
        All f.e. in pixel, inv. stage & 30.8 \tabularnewline
        Pre-amp. and driver in pixel & 23.4 \tabularnewline
        All f.e. in pixel & 27.1 \tabularnewline
        \hline 
    \end{tabular}
    \caption{RMS threshold dispersion $\sigma_{V_{th}}$ for each front-end configurations integrated in the chip.}
    \label{tab:sigma}
\end{table}
In addition, it is possible to notice that the pixels with the whole front-end integrated in the sensitive area are showing a higher threshold than the others. This effect is caused by the fact that, in this configuration, the PMOS transistors of the discriminator share the body with the pixel n-well, thus their threshold voltage is different from the one associated to PMOS integrated in external wells. 

The DACs integrated within the test-chip are based on a multiple current mirrors architecture in which the $i$-th input bit drives a mirror that doubles $i$ times a bias current (LSB). The variation of the LSBs showed in Figure \ref{LSB_disp} is caused by the mismatches of the converters. Because of the size of the DACs and their dispersion (a better matching would be obtained with bigger MOS transistors \cite{pelgrom1989matching}) this architecture will not be integrated in future prototypes of the FASER experiment. 
The final ASIC will feature either R-2R ladder DACs \cite{kennedy2000robustness} (characterized by a more compact architecture) or a set of converters placed in the periphery of the chip that will calibrate small sub-matrices. This choice is motivated by  the small threshold dispersion obtained in the test-chip and also by the significantly larger number of pixels that will be integrated in the final FASER ASIC. Moreover, the gain measurements reported in the following section confirm that the threshold dispersion is small enough to make the system able to meet the experiment requirement on the minimum input charge to discriminate (1 fC) even without using the local DACs. 
%The current mirror-based converters have been implemented on the presented chip only for test purposes and because they were already available from previous projects. 

%spiega plot due. Ricorda che si modifica la global th e si fa la media degli spostamenti dei codici
%th dispertion si calcola come vth=vdd_disc-(10+DAC)LSB. 10 è un offset.

\subsection{Tests with \textsuperscript{109}Cd source}

\begin{figure}[!t]
	\centering
	\subfloat[]{\includegraphics[width=2.5in]{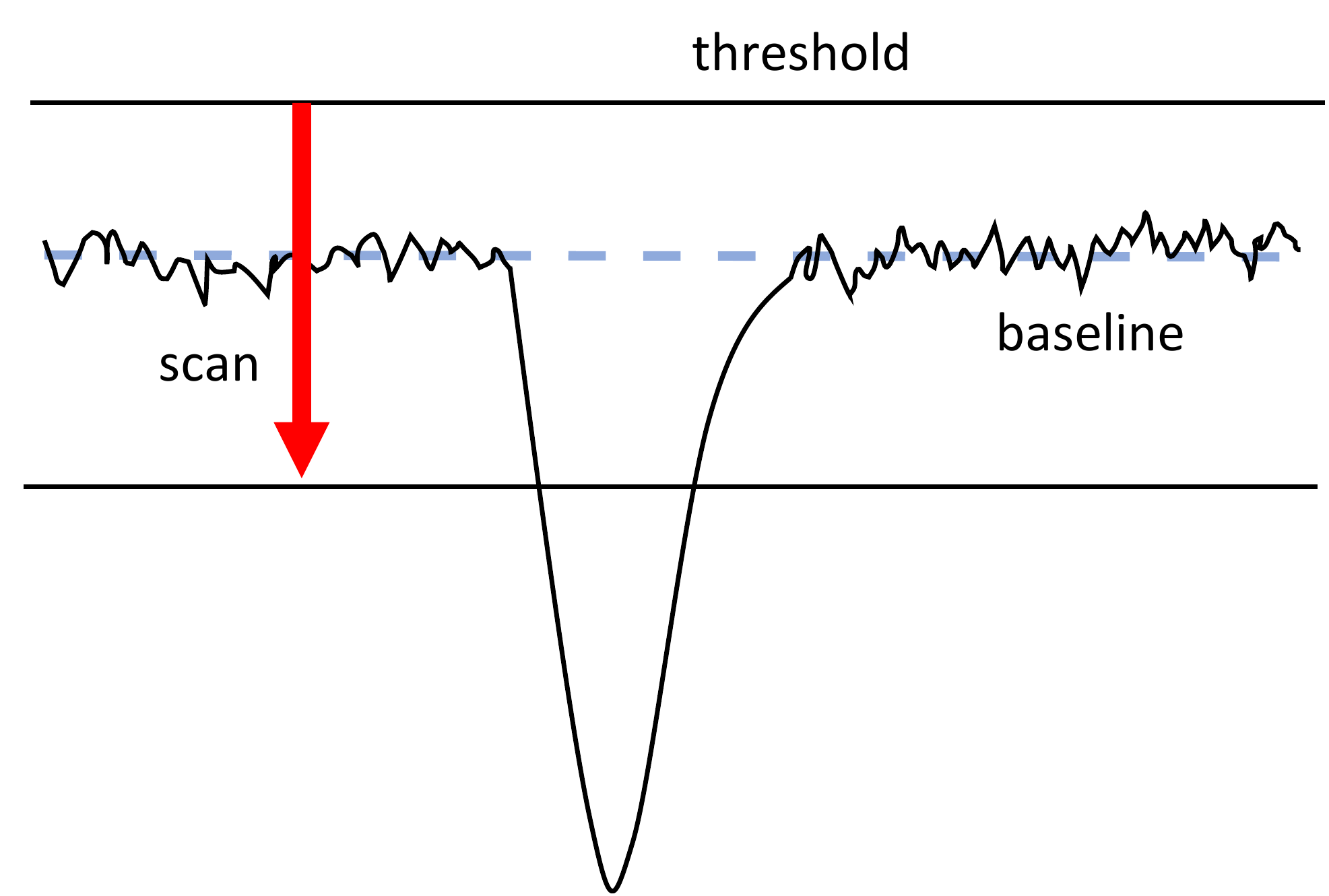}%
		\label{cadmium_scan}}
	\hfil
	\subfloat[]{\includegraphics[width=3.1in]{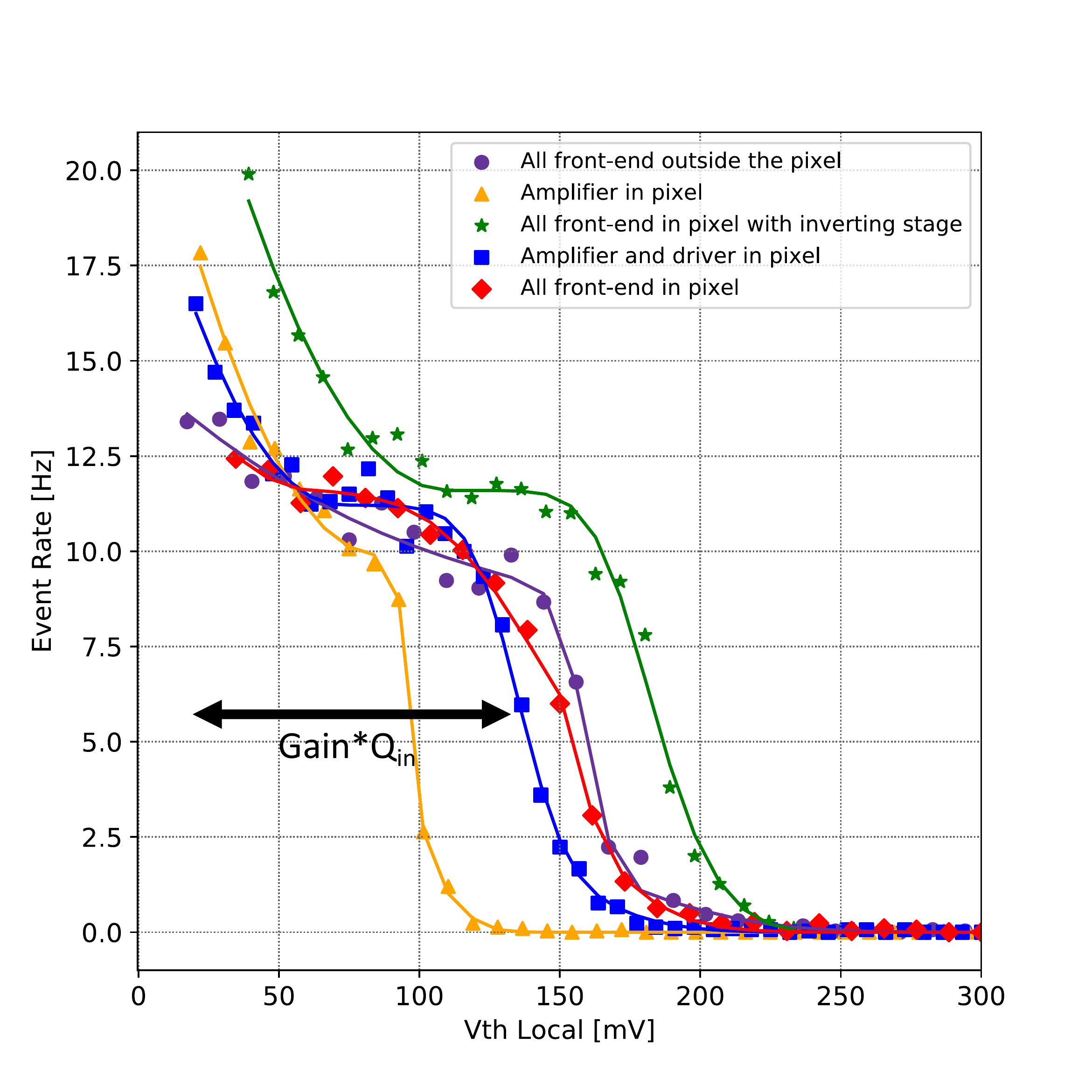}
		\label{cadmium_fit}}
	\caption{(a) Threshold scan representation for gain evaluation and (b) event rate as function of the threshold for different front-end configurations. The x-axis of (b) is referred to the baseline.}
	\label{Cadmium_scan_fit}
\end{figure}

A \textsuperscript{109}Cd radioactive source was used to measure the gain of the front-end circuits. The emission spectrum of this radioisotope is reported in \cite{cadmiumspectrum}. The gain evaluation was performed with a threshold scan as displayed in the representation of Figure \ref{cadmium_scan}: using the calibration data, the local threshold was set to an initial value close the baseline and then decreased (or increased, depending on the chosen front-end configuration). For each threshold step, the events were acquired and the event rate  recorded. Figure \ref{cadmium_fit} shows the result of one of such measurements. The data were then analysed using the  fitting function
\begin{subnumcases} {\label{eq:fitting_function} F_{Cd}(x)=}
N\cdot \mathrm{erfc}\left(\frac{x-\mu}{\sqrt{2}\sigma_1}\right)+0.28\cdot N\cdot \mathrm{erfc}\left(\frac{x-1.13\mu}{\sqrt{2}\sigma_2}\right) +a+bx+cx^2 & $x\leq \mu-2\sigma$\\
N\cdot \mathrm{erfc}\left(\frac{x-\mu}{\sqrt{2}\sigma_1}\right)+0.28\cdot N\cdot \mathrm{erfc}\left(\frac{x-1.13\mu}{\sqrt{2}\sigma_2}\right)+d & $x>\mu-2\sigma$
\end{subnumcases}
where $N$, $a$, $b$, $c$, $d$, $\sigma_{1,2}$ and $\mu$ are the fitting coefficients and $\mathrm{erfc}(x)$ is the complementary error function\footnote{The function  $\mathrm{erfc}(x)=1-\mathrm{erf}(x)=\frac{2}{\sqrt{\pi}}\int_{x}^{\infty}e^{-t^2}dt$ ~~where~~ $\mathrm{erf}(x)=\frac{2}{\sqrt{\pi}}\int_{0}^{x}e^{-t^2}dt$ ~~is the error function.}. The values 1.13 and 0.28 in the function are obtained evaluating the emission spectrum of the source and its peaks \cite{cadmiumspectrum}.  The polynomial is added to be able to fit the first part of the curve, related to the region in which the threshold is close to the baseline. The charge gain in mV/fC is equal to $\mu/0.98$ because the $\sim$22 keV photons emitted by the \textsuperscript{109}Cd source generate a ionization charge of approximately 0.98 fC in the 20-25 $\upmu$m depletion zone of this sensor.
\begin{figure}[!t]
	\centering
	\includegraphics[width=3.5in]{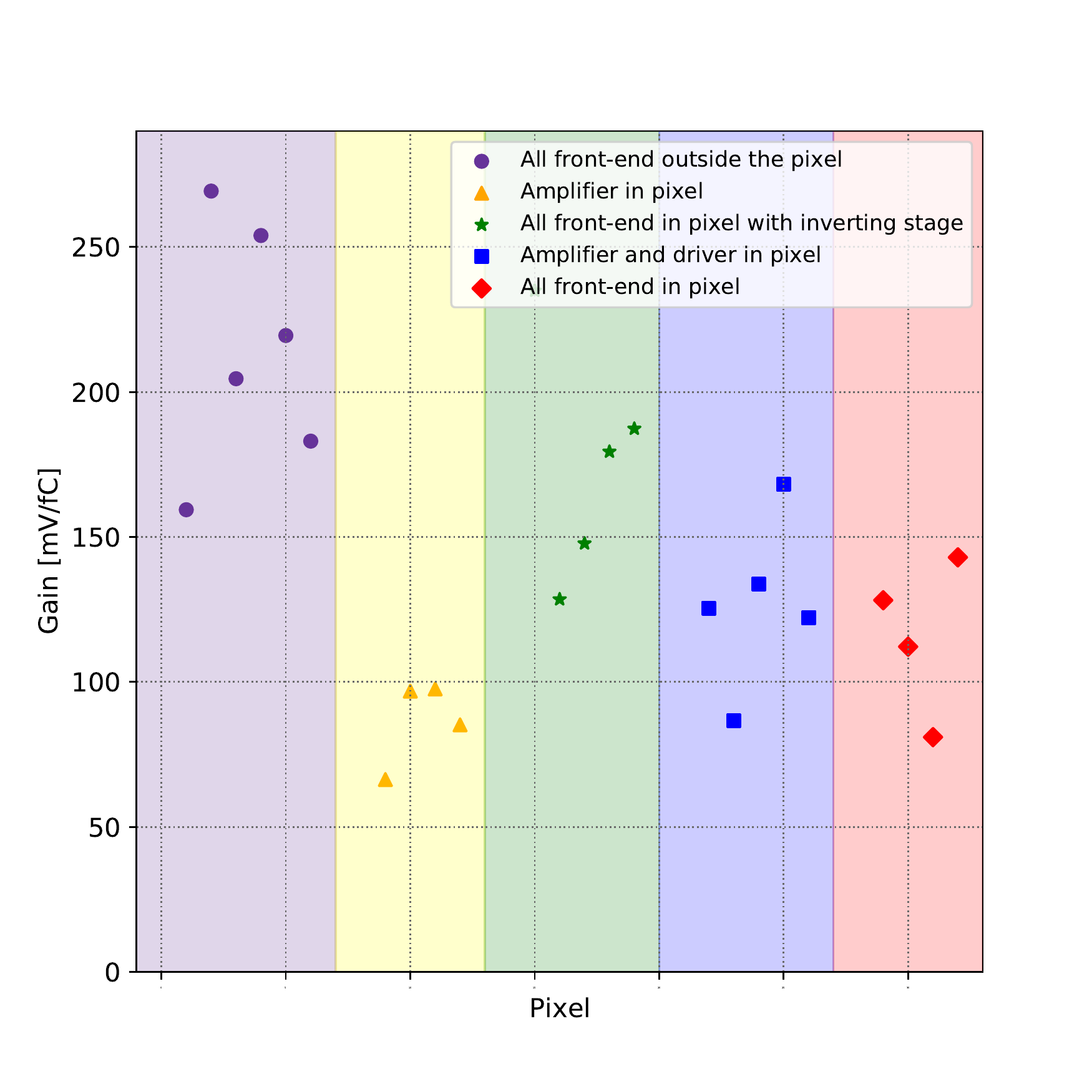}
	\caption{Gain of a selection of pixels that have been measured with the \textsuperscript{109}Cd source for the five front-end configurations.
	%\textcolor{red}{\\Fulvio, qui ed in figura 13 i punti delle differenti configurazioni sono colorati mentre in figura 12bottom non lo sono. Bisogna usare lo stesso stile... quindi dovresti mettere in colore i punti della figura 12}
	%\\La domanda che viene spontanea è: come è stata scelta questa selezione di pixels, e perchè sono in numero differente nelle diverse configurazioni?}
	}
	\label{gain_dispersion}
\end{figure}
Figure \ref{gain_dispersion} shows the value of the gain obtained with the above-mentioned method for some of the pixels for five of the front-end  configurations under study.

\subsection{Tests with \textsuperscript{55}Fe source}

A \textsuperscript{55}Fe radioactive source was used to measure the ENC associated to the front-end amplifiers. The \textsuperscript{55}Fe emission spectrum is characterized by a main peak at an energy of $\sim$5.9 keV \cite{ironspectrum}, that produces approximately 1650 e\textsuperscript{-} in the depletion region of our sensor. This charge is low enough to guarantee a linear response of the amplifiers.  In addition, the \textsuperscript{55}Fe peak is narrower than the 22 keV peak produced by the \textsuperscript{109}Cd radioisotope that generates an input charge of approximately 1 fC $\approx$ 6240 e\textsuperscript{-}. Because of the narrower peak and lower energy, the \textsuperscript{55}Fe source is more suited to analyse the noise contribution of the front-end system and was used for our measurements. 

\begin{figure}[!t]
	\centering
	\includegraphics[width=4in]{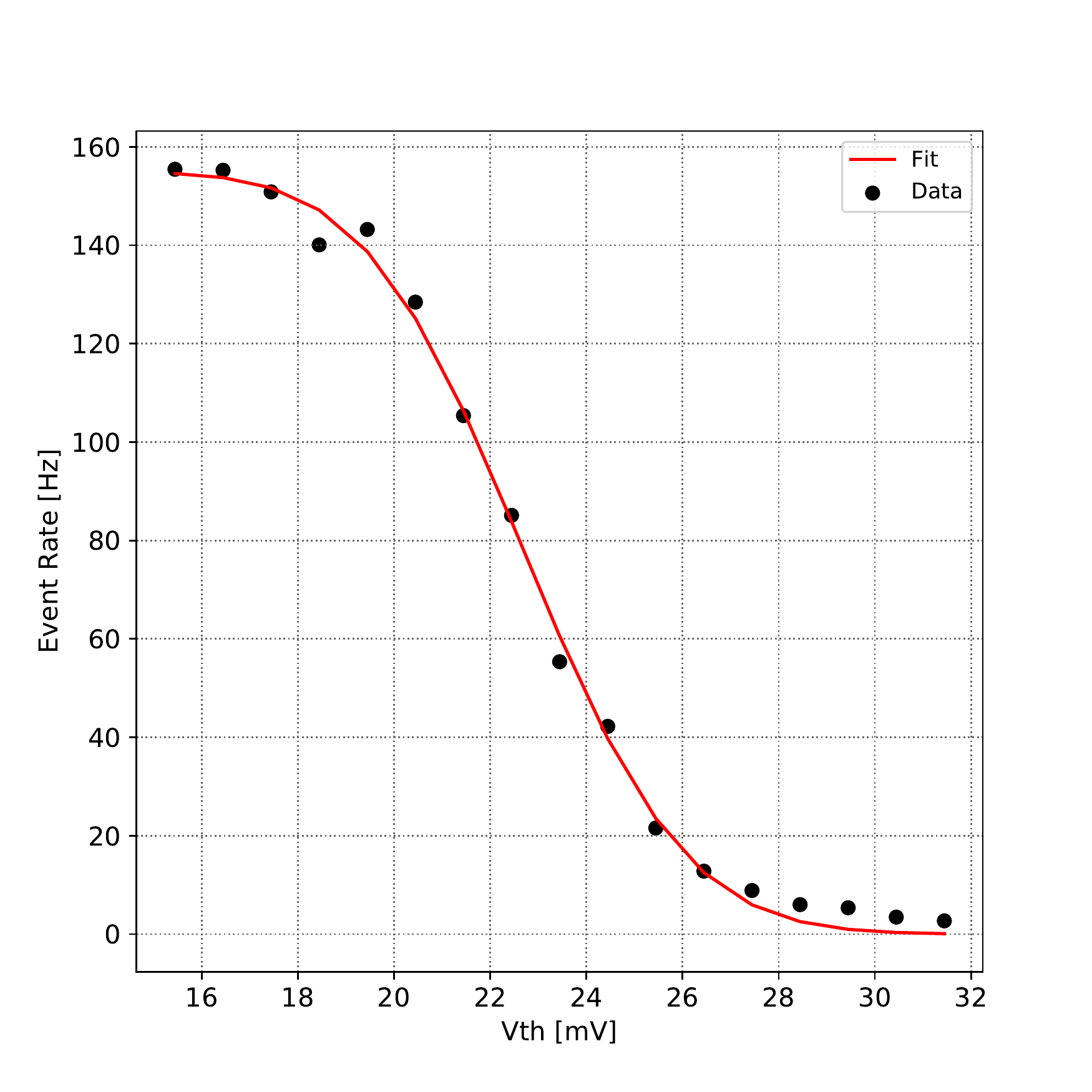}
	\caption{Event rate as a function of the threshold obtained with a \textsuperscript{55}Fe radioactive source.}
	\label{iron_fit}
\end{figure}

A threshold scan was performed with the \textsuperscript{55}Fe source and the event-rate data were fitted with the  function
\begin{equation}
\label{eq:fitting_function_fe}
    F_{Fe}(x)=N\cdot \mathrm{erfc}\left(\frac{x-\mu}{\sqrt{2}\sigma_v}\right)+9.3\cdot N\cdot \mathrm{erfc}\left(\frac{x-0.9\mu}{\sqrt{2}\sigma_v}\right)
\end{equation}
to calculate the $\sigma_v$ component of Equation \ref{eq:sigma_elec} and from it obtaining the equivalent-noise charge as 
%component of Equation \ref{eq:sigma_elec} 
\begin{equation}
    ENC=\frac{\sigma_v}{G_c},
\end{equation}
%~~\textcolor{red}{{Fulvio: NON C'E` NESSUNA SIGMAV NELLA FORMULA 3.2}} 
where $G_c$ is the charge gain of the amplifier measured with the \textsuperscript{109}Cd source. 
%The value of $\sigma_v$ is directly obtained by the $\sigma$ of the fitting function $F_{Fe}(x)$: a noisier amplifier would produce a smoother event rate function. 
Figure \ref{iron_fit} shows the event rate distribution as a function of the threshold obtained with the \textsuperscript{55}Fe source for one of the pixels of the matrix. 
%In the text section the front-end performance will be compared and discussed.

\subsection{Analysis of the performance and configurations comparison}
\label{sec:comparison}

\begin{table}[!t]
    \centering
    \begin{tabular}{lccc}
        \hline 
        Configuration & $\sigma_v$ [mV] & $G_c$ [mV/fC] & $ENC$ [e\textsuperscript{-}]\tabularnewline
        \hline 
        All f.e. outside pixel & 4.2 $\pm$ 0.2          & ~~~159 $\pm$ 1.0       & 165 $\pm$ 9 \tabularnewline
        Only pre-amp. in pixel & 2.5 $\pm$ 0.1          & ~~96.8 $\pm$ 0.5        & 161 $\pm$ 9  \tabularnewline
        All f.e. in pixel, inv. stage & 6.9 $\pm$ 0.5   & ~~~179 $\pm$ 1.0       & ~~241 $\pm$ 19\tabularnewline
        Pre-amp. and driver in pixel~~~ & 3.8 $\pm$ 0.2    & 133.7 $\pm$ 0.6       & 178 $\pm$  9 \tabularnewline
        All f.e. in pixel & 5.4 $\pm$ 0.4               & ~~~148 $\pm$ 1.0       & ~~228 $\pm$ 20 \tabularnewline
        \hline 
    \end{tabular}
    \caption{Noise contribution ($\sigma_v$), charge gain ($G_c$), ENC of one channel for each front-end configurations integrated in the chip and threshold dispersion ($\sigma_{V_{th}}$). The error associated to $G_c$ does not represent the channel-to-channel dispersion but is the uncertainty on the gain measurement of the analysed channel.}
    \label{tab:enc_gain_sigma}
\end{table}

Table \ref{tab:enc_gain_sigma} reports a summary of the measurements performed on one channel for each front-end configurations integrated in the ASIC. 
The configuration in which the front-end system is completely outside the pixel is the one that shows the best performance in terms of noise. Integrating the electronics outside the sensitive area is useful to reduce the noise on the pixel induced by the amplifiers. The gain of the architecture is one of the highest among all because, despite the connection with the pre-amplifier input is longer than the other versions (in which at least the first stage of the front-end is inside the pixel), the capacitance of the sensitive area is smaller since no triple-well is needed for the integration of the electronics. However, this solution is the one that requires using the most area outside the pixel and it is vulnerable to the scaling of the pixel density and, as anticipated in Section \ref{sec:flavours}, it will not be adopted in the final FASER chip.

Integrating only the pre-amplifier inside the pixel leads to a significant reduction of the charge gain. This effect is caused by the need of a longer connection between the pre-amplifier and the driver and consequently of the increase of the output capacitance of the former which results into a reduction of its bandwidth. Moreover, a longer pre-amplifier output line increases the coupling with the pixel in which the circuit is integrated. The first stage of the front-end is inverting the pixel signal leading to a negative feedback with the sensor. Therefore, a more intense coupling (due to the longer lines) is further reducing the charge gain.

The configurations with a third (inverting) stage (depicted in red in Figure \ref{sec:pre-amp}) is characterized by the largest measured gain. The additional block is improving the decoupling between the first driver and the discriminator resulting into an increase of $G_c$. However, a second driving stage worsens the noise performance of this solution which makes it the worst in terms of ENC among all. 

The front-end variants that include every stage inside the sensitive area of the pixel are characterized by the second worst ENC performance. Also, the charge gain is not the highest among the designed configurations because, as explained before, the triple-well in which the circuits are integrated increases the sensor capacitance. This explains also the noise performance. Similar performance in terms of gain but better ENC are achieved by the configurations with only pre-amplifier and driver in pixel. In this case the exclusion of the discriminator inside the sensitive pixel area  significantly improves the noise level of the front-end and reduces the ENC. The last two solutions represent a good compromise between performance and compactness, a crucial requirement for the design of the final version of the FASER pre-shower chip. For this reason, they will be taken into consideration for the next iterations.

Table \ref{tab:sigma} showed that all the solutions are characterized by a threshold dispersion $\sim$30 mV, or smaller, in some configurations.
In particular, the variant with the discriminator outside the pixel  shows a peak-to-peak dispersion of $6\cdot\sigma_{V_{th}}=140.4$ mV and an average gain  around 130 mV/fC. This front-end configuration will therefore  meet the specification of the high-precision FASER pre-shower to be able to discriminate input charges of $Q_{in}\gtrsim1$ fC even without using local DACs for pixel-to-pixel calibration. As anticipated, this solution will be investigated for future implementations of the chip.
%This simplified system would meet the specification of the high-precision pre-shower of the FASER experiment.
%Math mode: $\SI{50 \pm 2}{\percent}$.

%1) all f.e. inside: non ha il miglior guadagno perche la capacita in pixel aumenta all'aumentare dei circuiti integrati in pixel. peggior rumore proprio perche ci sono piu componenti. 

%2) pre-amp and driver in pixel: guadagno simile ma meno rumore -> assenza del discriminatore in pixel. Soluzione che verra implementata in FASER

%3) inverting stage: disegnato per avere il guadagno piu alto. ha anche il secondo peggior rumore per l'aumento del numero di componenti

%4) only pre-amp in pixel: guadagno piu basso per l'alta capacita tra driver e pre-amp. inoltre l'uscita del pre-amp fa feedback negativo con il pixel e la linea ci passa sopra. 

%5) all f.e. outside pixel: miglior rumore perche tutto sta fuori. non implementabile nel chip finale per l'alta densita di componenti e di linee che ne ridurrebbe il guadagno. 

%migliori compromessi 4) e 2) ma si opterebbe per 2)

\section{Conclusions}

A small prototype  was designed to evaluate the performance of sevearal front-end configurations for the monolithic ASIC of the high-resolution pre-shower upgrade of the FASER experiment at CERN. 
The analysis of these circuits was crucial to choose the most suitable way to integrate the front-end system inside the final version of the  ASIC. 
The stability of the system, optimised with dedicated design features with the support of Cadence Spectre simulations, was confirmed by the measurement results. 
The tests showed that a careful layout design could prevent the onset of unwanted oscillation and instabilities, common problems for fast front-end amplifiers in pixels. 
Different degrees of integration of the front-end in the pixel were studied. The measurements showed that the configuration with pre-amplifier and driver in the sensitive area has a charge gain above 130 mV/fC and ENC of $\sim$180 e\textsuperscript{-}. 
The configuration with the highest degree of in-pixel integration, which has also the discriminator in pixel, shows similar performance in terms of gain and an ENC of $\sim$230 e\textsuperscript{-}. 
Both these results are compatible with the requirements of the final FASER pre-shower ASIC and will be taken into consideration for further developments. 
The pixel threshold calibration circuit, implemented specifically for this prototype and not foreseen for future iterations, showed poor performance in terms of mismatch, but the measurement of the intrinsic threshold dispersion of the front-end confirms that it is possible to set a threshold of $\gtrsim$1 fC without the need for a pixel-by-pixel calibration. 

\acknowledgments

The authors wish to thank Y. Favre, S. Dèbieux and all the technical staff of the Particle Physics Department (DPNC) at University of Geneva for their contribution to the design of the boards used for the tests of the ASICs.
This research is funded by the Swiss National Science Foundation grants  200020-188489 and 20FL21-201474.

%\paragraph{Note added.} Text note

\bibliographystyle{unsrt}
\bibliography{mybib}

% \begin{thebibliography}{99}

% \bibitem{a}
% Author, \emph{Title}, \emph{J. Abbrev.} {\bf vol} (year) pg.

% \bibitem{b}
% Author, \emph{Title},
% arxiv:1234.5678.

% \bibitem{c}
% Author, \emph{Title},
% Publisher (year).

% % Please avoid comments such as "For a review'', "For some examples",
% % "and references therein" or move them in the text. In general,
% % please leave only references in the bibliography and move all
% % accessory text in footnotes.

% % Also, please have only one work for each \bibitem.

% \end{thebibliography}
\end{document}